\documentclass[twocolumn]{aastex62}
\usepackage{lineno}
\usepackage{amsmath}
\usepackage{lipsum}
\usepackage{blindtext}
\usepackage{graphicx} 
\usepackage{changepage}
\usepackage{array}
\usepackage{multirow}
\usepackage{tabularx}

\begin{document}
	\title{\textbf{A Catalog of 71 Coronal Line Galaxies in MaNGA: [NeV] is an Effective AGN Tracer}}
	\shorttitle{Coronal Lines in MaNGA
	}
	\shortauthors{Negus et al.}
	
	\author[0000-0003-2667-7645]{James Negus}
	\email{$^{\star}$ james.negus@colorado.edu}
	\affil{The University of Colorado Boulder, 2000 Colorado Avenue, Boulder, CO 80309, USA}
	
	\author[0000-0001-8627-4907]{Julia M. Comerford}
	\affil{The University of Colorado Boulder, 2000 Colorado Avenue, Boulder, CO 80309, USA}
	
	\author[0000-0002-2713-0628]{Francisco M\"uller S\'anchez}
	\affiliation{The University of Memphis, 3720 Alumni Avenue, Memphis, TN 38152, USA}
	
	\author[0000-0002-4917-7873]{Mitchell Revalski}
	\affiliation{Space Telescope Science Institute, 3700 San Martin Drive, Baltimore, MD 21218, USA}
	
	\author[0000-0003-0483-3723]{Rogemar A. Riffel}
	\affiliation{Departamento de Física, CCNE, Universidade Federal de Santa Maria, 97105-900, Santa Maria, RS, Brazil}
	\affiliation{Laboratório Interinstitucional de e-Astronomia - LIneA, Rua Gal. José Cristino 77, Rio de Janeiro, RJ - 20921-400, Brazil}
	
	\author[0000-0001-9742-3138]{Kevin Bundy}
	\affiliation{UC Santa Cruz, 1156 High Street, Santa Cruz, CA 95064}
	
	\author[0000-0003-1056-8401]{Rebecca Nevin}
	\affiliation{Fermi National Accelerator Laboratory, Batavia, IL 60510, USA}
	
	\author[0000-0003-0880-5738]{Sandro B. Rembold}
	\affiliation{Departamento de Física, CCNE, Universidade Federal de Santa Maria, 97105-900, Santa Maria, RS, Brazil}
	\affiliation{Laboratório Interinstitucional de e-Astronomia - LIneA, Rua Gal. José Cristino 77, Rio de Janeiro, RJ - 20921-400, Brazil}
	

\begin{abstract}
Despite the importance of AGN in galaxy evolution, accurate AGN identification is often challenging, as common AGN diagnostics can be confused by contributions from star formation and other effects (e.g., Baldwin-Phillips-Terlevich diagrams). However, one promising avenue for identifying AGNs are “coronal emission lines” (``CLs’’), which are highly ionized species of gas with ionization potentials $\ge$ 100 eV. These CLs may serve as excellent signatures for the strong ionizing continuum of AGN. To determine if CLs are in fact strong AGN tracers, we assemble and analyze the largest catalog of optical CL galaxies using the Sloan Digital Sky Survey’s Mapping Nearby Galaxies at Apache Point Observatory (MaNGA) catalog. We detect CL emission in 71 MaNGA galaxies, out of the 10,010 unique galaxies from the final MaNGA catalog, with $\ge$ 5$\sigma$ conﬁdence. In our sample, we measure  [NeV]$\lambda 3347$, $\lambda 3427$, [FeVII]$\lambda 3586$, $\lambda 3760$, $\lambda 6086$, and [FeX]$\lambda 6374$ emission and crossmatch the CL galaxies with a catalog of AGNs that were confirmed with broad line, X-ray, IR, and radio observations. We find that [NeV] emission, compared to [FeVII] and [FeX] emission, is best at identifying high luminosity AGN. Moreover, we find that the CL galaxies with the least dust extinction yield the most iron CL detections. We posit that the bulk of the iron CLs are destroyed by dust grains in the galaxies with the highest [OIII] luminosities in our sample, and that AGN in the galaxies with low [OIII] luminosities are possibly too weak to be detected using traditional techniques. 



\end{abstract}
\section{Introduction} \label{sec:intro}
Active Galactic Nucleus (AGN) feedback, the process by which an active accretion disk converts gravitational energy into radiative or mechanical energy (e.g., AGN-induced photoionization, outflows, shocks, winds, and jets), has been shown to dynamically influence the evolution of a host galaxy (e.g., the tight correlation between stellar velocity dispersion and black hole mass and the quenching of star formation; e.g., \citealt{2000ApJ...539L...9F, 2000ApJ...539L..13G,2005ApJ...630..705H,2005Natur.433..604D, 2012ARA&A..50..455F, 2013ARA&A..51..511K, 2014ARA&A..52..589H}). However, the full spatial extent, ionization properties, and impact of AGN feedback on the host galaxy have yet to be fully unraveled. 

The Unified Model of AGN \citep{1993ARA&A..31..473A, 1995PASP..107..803U} provides a fundamental architecture for  understanding the evolution of AGN feedback. In this model, an AGN is either Type I or Type II. Type I are viewed pole-on and are observed to have broad (FWHM $\gtrapprox$ 1,000 km s$^{-1}$) and narrow (FWHM $\lessapprox$ 1,000 km s$^{-1}$) emission lines, whereas Type II  are viewed edge-on and are observed to only have narrow emission lines. These regions are termed the broad-line region (BLR) and the narrow-line region (NLR), respectively.

In \citealt{2021ApJ...920...62N}, we considered the Unified Model before investigating the ``coronal line region" (CLR), an area surrounding a supermassive black hole (SMBH; M$_{\rm{BH}} > 10^{6}$ M$_{\odot}$) that produces highly ionized species of gas with ionization potentials (IPs) $\gtrsim$ 100 eV (termed ``coronal lines" (CLs) since they were first observed in the solar corona). CLs are suspected to primarily originate from the strong ionizing continuum of an AGN; in particular, nuclear CLs are produced in the inner edge of the dusty torus and extended CLs are tied to the presence of a jet or AGN-driven outflows (due to the highly energetic nature of these processes; e.g., \citealt{2002ApJ...579..214R, 2005MNRAS.364L..28P, 2009MNRAS.397..172G,2009MNRAS.394L..16M,2010MNRAS.405.1315M,2011ApJ...743..100R, 2011ApJ...739...69M, 2016ApJ...824...34G, 2021MNRAS.503.5161R, 2022MNRAS.511.1420T}).

 Further, CLs in the mid-infrared have been extensively used to probe for AGNs, and to subsequently analyze their physical environments, within dusty galaxies (e.g., \citealt{1998ApJ...498..579G}, \citealt{2002A&A...393..821S},
\citealt{2004ApJS..154..178A} \citealt{2005ApJ...632L..13L},
\citealt{2006ApJ...651..101W}, \citealt{2008ApJ...674L...9D}, \citealt{2022arXiv220913125A}). In fact, several studies have shown that AGNs, even those missed by optical surveys (due to obscuration, for example), are uncovered by observations of infrared CLs (e.g., \citealt{2008ApJ...677..926S,2021ApJ...906...35S,2022Univ....8..356S}). Additionally, since CL emission from Type II supernovae is infrequent, weak, and short lived, CL infrared observations have been particularly useful for accurately identifying CL emission exclusively from AGNs (e.g., \citealt{2009ApJ...695.1334S}).

In regard to optical studies, Baldwin-Phillips-Terlevich diagnostics diagrams (\citealt{1981PASP...93....5B,1987ApJS...63..295V, 2001ApJ...556..121K, 2006MNRAS.372..961K}) are predominantly used to differentiate emission-line sources as star-forming, AGN, or a composite of the two. However, diffuse ionized gas, extraplanar gas, photoionization by hot stars, metallicity, and shocks can elevate sources beyond the star formation threshold and potentially lead to AGN misclassification (e.g., \citealt{2018MNRAS.474.1499W}). 

Moreover, while the NLR is the largest observable structure directly affected by an AGN's ionizing radiation (out to several kpcs; e.g., \citealt{2011ApJ...739...69M}), star formation can also produce some of the narrow lines usually associated with AGN (e.g., [OIII] 5007; ``[OIII]" hereafter). Further, while the BLR provides definitive evidence of AGN activity, due to the elevated cloud velocities, its compact radial extent ($\approx$ 0.1 kpc; e.g., \citealt{2004ASPC..311..169L}) is spatially unresolved in most spectroscopic surveys.  On the other hand, CLs require energies well above the limit of stellar emission (55 eV; \citealt{2001ApJ...549L.151H}) and are typically spatially resolved beyond the BLR and well into the NLR (e.g., \citealt{2021ApJ...920...62N}).  If CLs can provide accurate AGN identification in optical spectroscopic surveys of galaxies, as they have been shown to do in infrared surveys, then detecting them may be a critical step in constraining the complexities of AGN feedback (e.g., \citealt{2021ApJ...922..155M}).

The Sloan Digital Sky Survey's (SDSS) Mapping Nearby Galaxies at Apache Point Observatory catalog (MaNGA; \citealt{2015ApJ...798....7B}) has provided an unprecedented lens into the dynamic environments that surround the SMBHs of nearly 10,010 nearby (0.01 $<$ \textit{z} $<$ 0.15; average \textit{z} $\approx$ 0.03) galaxies. Using integral field spectroscopy (IFS), MaNGA provides a 1 - 2 kpc spatial sampling across the field of view of each observed galaxy, which offers direct insight into the spatial extent, ionization properties, and the environmental impact of AGN feedback. For reference, previous SDSS surveys (e.g., SDSS-I to SDSS-III; \citealt{2000AJ....120.1579Y, 2011AJ....142...72E}) observed galaxies with small (3'' diameter) optical fibers. The resulting spectra only traced a small region close to the galactic center, potentially missing nuclear activity outside of this region. \cite{2016AJ....152..197Y} further report that 80\% of SDSS galaxies observed with a single fiber have less than 36\% of their light covered. Moreover, long-slit spectroscopic surveys of galaxies also reveal limited spatial information, since only narrow elongated regions of each galaxy are observed (e.g., \citealt{2013ApJS..208....5N}). In contrast, MaNGA  offers the ability to capture spatially extended galactic features, which can reveal off-nuclear activity and large-scale emission line regions.
%

In \citealt{2021ApJ...920...62N}, we scanned for [NeV]$\lambda 3347$, $\lambda 3427$, [FeVII]$\lambda 3586$, $\lambda 3760$, $\lambda 6086$, and [FeX]$\lambda 6374$ emission at $\ge 5 \sigma$ above the background continuum in the 6,623 galaxies from MaNGA's eighth data release (MPL-8). We identified 10 CL galaxies in MPL-8, the largest such catalog at the time; seven of which were confirmed to host an AGN, which suggests that CL emission can be useful for tracing AGN activity. The remaining three visually appear to be undergoing galactic mergers. We also found that the average spatial extent of the CLR from the nuclear center is 6.6 kpc - well into the NLR. Further, we measured the average electron number density of the CLRs in our sample to be on the order of $\approx$ 10$^{2}$ cm$^{-3}$, also consistent with the CLR occupying the traditional NLR, beyond the BLR (typical NLR densities range from 10$^{1}$ - 10$^{7}$ cm$^{-3}$; e.g.,  \citealt{1997iagn.book.....P, 2022ApJ...930...14R}). 

However, we also reported a range of power-law indices ($\alpha$) above the threshold expected for pure AGN photoionization ($\alpha$ = $-2$; we measured -1.8 ± 0.3 $\le$ $\alpha$ $\le$ 0.2 ± 0.1), and electron temperature values slightly above the threshold for pure AGN photoionization (T$_{e} = 20,000$ K;  \citealt{1981ApJ...246..696O}). We found that the average CLR electron temperatures varied between 12,331 K - 22,530 K. These results suggest that shock-induced compression and heating may also play a role in the production of CLs. 

Comparatively,  \cite{2010MNRAS.405.1315M} investigated the CLR for 10 pre-selected AGNs. They used the  \textit{Hubble Space Telescope}/Space Telescope Imaging Spectrograph to study [NeV] $\lambda$3427, [FeVII] $\lambda$3586, $\lambda$3760, $\lambda$6086, [FeX] $\lambda$6374, [FeXIV] $\lambda$5303, [FeXI] $\lambda7892$, and [SXII] $\lambda$7611 emission in their sample. The authors deduced that AGN photoionization is the main driving mechanism for the CLs.  Moreover, \cite{2009MNRAS.397..172G} used the sixth SDSS data release (\citealt{2008ApJS..175..297A}) to analyze the CLR in 63 AGNs with [FeX] $\lambda$6374 (IP = 233.60 eV), [FeXI] $\lambda$7892 (IP = 262.10 eV), and [FeVII] $\lambda$6086 (IP = 99.10 eV) emission. They used X-ray observations from \textit{Rosat} (\citealt{1999A&A...349..389V,2000IAUC.7432....3V}) to similarly posit that AGN photoionization is the main ionization source of the CLs. Finally, \cite{2022ApJ...936..140R} executed the first systematic survey of twenty optical CLs in the spectra of nearly 1 million galaxies from the eighth SDSS data release (\citealt{2011ApJS..193...29A}). The authors found that CL emission is extremely rare ($\approx 0.03\%$ of the sample show at least one CL), and that the highest ionization potential CLs tend to be found in lower mass galaxies. They reasoned that this finding is consistent with theory that hotter accretion disks are produced by lower mass black holes, which typically reside in lower mass galaxies. 

Here, we use MaNGA's eleventh, and final, data release (MPL-11; 10,010 unique galaxies) to further resolve the physics of the CLR, and to better understand the relationship between the production of CLs and AGN activity. With our custom pipeline, we identify 71 unique galaxies with emission from either [NeV]$\lambda$3347, $\lambda$3427, [FeVII]$\lambda$3586, $\lambda$3760, $\lambda$6086, or [FeX]$\lambda$6374 detected at $\ge 5\sigma$ above the continuum, which makes it the most extensive such catalog of MaNGA CL galaxies to date.

This paper is outlined as follows: Section \ref{sec:obs} details the technical components of the SDSS-IV MaNGA survey and its data pipeline, Section \ref{sec:analysis} describes the methodology we use to build the CL catalog and to analyze the physical properties of the CLR, Section \ref{sec:results} reviews our results,  Section \ref{sec:discussion} provides interpretations of our findings, and Section \ref{sec:conclusion} includes our conclusions and intended future work. All wavelengths are provided in vacuum and we assume a $\Lambda$CDM cosmology with the following values: $\Omega_{M} = 0.287$, $\Omega_{\Lambda} = 0.713$ and $H_{0} = 69.3$ $ \rm{km}$ $\rm{s}^{-1}$ $\rm{Mpc}^{-1}$.
\section{Observations}
\label{sec:obs}
\subsection{Sample of Galaxies}
We assemble our sample from the SDSS-IV MaNGA catalog \citep{2015ApJ...798....7B,2015AJ....149...77D,2016AJ....152...83L, 2016AJ....152..197Y, 2017AJ....154...28B, 2017AJ....154...86W}. MaNGA observations occurred between 2014 to 2020, using the SDSS 2.5 m telescope \citep{2006AJ....131.2332G}. The IFS survey contains data for 10,010 nearby galaxies (0.01 $<$ \textit{z} $<$ 0.15; average \textit{z} $\approx$ 0.03) with stellar mass distributions between $10^{9}$ $M_{\odot}$ and $10^{12}$ $M_{\odot}$. The spectra were taken at wavelengths between 3622 \AA$\>$ - 10354 \AA, with a typical spectral resolving power of $\approx$ 2000, corresponding to a velocity resolution of $\approx$ 60 km s$^{-1}$ (see \citealt{2015ApJ...798....7B}). 

MaNGA contains spectroscopic maps out to at least 1.5 times the effective radius; the typical galaxy is mapped out to a radius of 15 kpc. Each MaNGA spatial pixel, or spaxel, covers 0.$^{\prime\prime}$5 $\times$ 0.$^{\prime\prime}$5, and the average full-width half maximum (FWHM) of the on-sky point spread function (PSF) is 2.$^{\prime\prime}$5, which corresponds to a typical spatial resolution of 1 -2  kpc \citep{2015AJ....149...77D}.  

\subsection {MaNGA Data Analysis Pipeline}

The MaNGA Data Analysis Pipeline (DAP; \citealt{2019AJ....158..231W}) offers publicly available high-level data products. The MaNGA DAP algorithms have been in development since 2014 and its main outputs are stellar kinematics, fluxes and kinematics of prominent emission lines, and continuum spectral indices. To measure each parameter, the DAP relies on spectral fitting with pPXF (\citealt{2012ascl.soft10002C,2017MNRAS.466..798C}), where each fit features a blend of stellar templates with a multiplicative polynomial component to the stellar continuum. In particular, the DAP incorporates the MILESHC stellar templates library (\citealt{2019AJ....158..231W}) to fit the stellar kinematics. 

The inputs for the DAP are data reduced by the MaNGA Data Reduction Pipeline (DRP). The DRP is fed spectra from the MaNGA fiber-feed system, which consists of 17 IFUs: two 19-fiber IFUs, four 37-fiber IFUs, four 61-fiber IFUs, two 91-fiber IFUs, and five 127-fiber IFUs (see \citealt{2015AJ....149...77D} for a more detailed description). The DRP subsequently wavelength, flux, and astrometrically calibrates the spectra. 
\section{Analysis}
\label{sec:analysis}

\subsection {CL Continuum Subtraction and Emission Line Fitting}

We scan for  [NeV]$\lambda 3347$, $\lambda 3427$, [FeVII]$\lambda 3586$, $\lambda 3760$, $\lambda 6086$, and [FeX]$\lambda 6374$ emission to better understand their effectiveness as AGN indicators. These CLs are selected because MaNGA’s DAP does not provide emission line measurements for them. As a result, we expand upon the custom pipeline detailed in \citealt{2021ApJ...920...62N} to measure these CLs in MPL-11. Note, all 10 CL galaxies reported in \citealt{2021ApJ...920...62N} are recovered using the new MPL-11 pipeline. 

\subsubsection {CL Stellar Continuum Subtraction}
\label{cont}

To measure the stellar kinematics, and subsequently subtract the stellar continuum for each CL galaxy’s observed spectra, we use pPXF (\citealt{2012ascl.soft10002C,2017MNRAS.466..798C}). pPXF performs a polynomial fit on each galaxy’s spectrum while masking gas emission lines. For each fit, we use the MILES\footnote{\href{http://miles.iac.es/}{http://miles.iac.es/}} stellar templates library to represent the stellar population synthesis model. This library contains $\approx$ 1,000 stars, with spectra obtained by the \textit{Isaac Newton Telescope}. These spectra cover the wavelength range of 3525 \AA $\>$- 7500 \AA $\>$at a 2.5 \AA $\>$FWHM resolution. 

We ﬁrst access the DRP to extract the necessary data cubes for each MaNGA galaxy before performing the pPXF stellar continuum subtraction. The data cubes provide a spectrum for each individual spaxel across the FoV of each galaxy. We then use the spectroscopic redshifts of each galaxy, adopted from the NASA Sloan Atlas catalogs (\citealt{2011AJ....142...31B}), to adjust the spectra to rest vacuum wavelengths. We also use a minimum redshift threshold (\textit{z}$_{\rm{min}}$) for CLs near the lower wavelength limit of MaNGA (3622 \AA; Table \ref{tab:coronal}) to ensure CLs of interest are not shifted out of MaNGA's spectral coverage. For [NeV]$\lambda$3347, $\lambda$3427 and [FeVII]$\lambda$3586, $\approx$ 93\% (9,152), $\approx$ 83\% (8,096), and $\approx$ 3\% (229) of the MPL-11 galaxies, respectively, feature redshifts that place each CL out of MaNGA’s spectral range; as a result, we are unable to scan for [NeV]$\lambda$3347, $\lambda$3427 and [FeVII]$\lambda$3586 in these respective galaxies.

We then apply a mask to each datacube, such that the imported wavelength range for each spectrum matches the wavelength range of the stellar templates library (3525 \AA - 7500 \AA). Next, we normalize each spectrum by dividing fluxes in this wavelength range by each spectrum's median flux value (to avoid numerical issues; see \citealt{2017MNRAS.466..798C} for a more detailed discussion). Subsequently, we define a typical instrument resolution of $\approx$ 2.5 \AA, construct a set of Gaussian emission line templates (to mask emission lines; provided by pPXF), and fit the stellar templates. Note, for the CLs near the lower limit of the mask (3525 \AA; e.g., [NeV]$\lambda$3347, $\lambda$3427 and [FeVII]$\lambda$3586), we perform a custom stellar continuum fit and subtraction before measuring the target emission line. In these instances, we execute a polynomial ﬁt on a narrow spectral region, $\approx$ 300 \AA $\>$ wide, of continuum (free of prominent absorption or emission lines) near the rest wavelength of the target CL to model the background stellar continuum and subtract it from the spectrum. 

\subsubsection {[NeV] and [FeVII] Emission Line Measurements}
\label{emline}

\begin{figure}[t]
	\includegraphics[height = 2.1in]{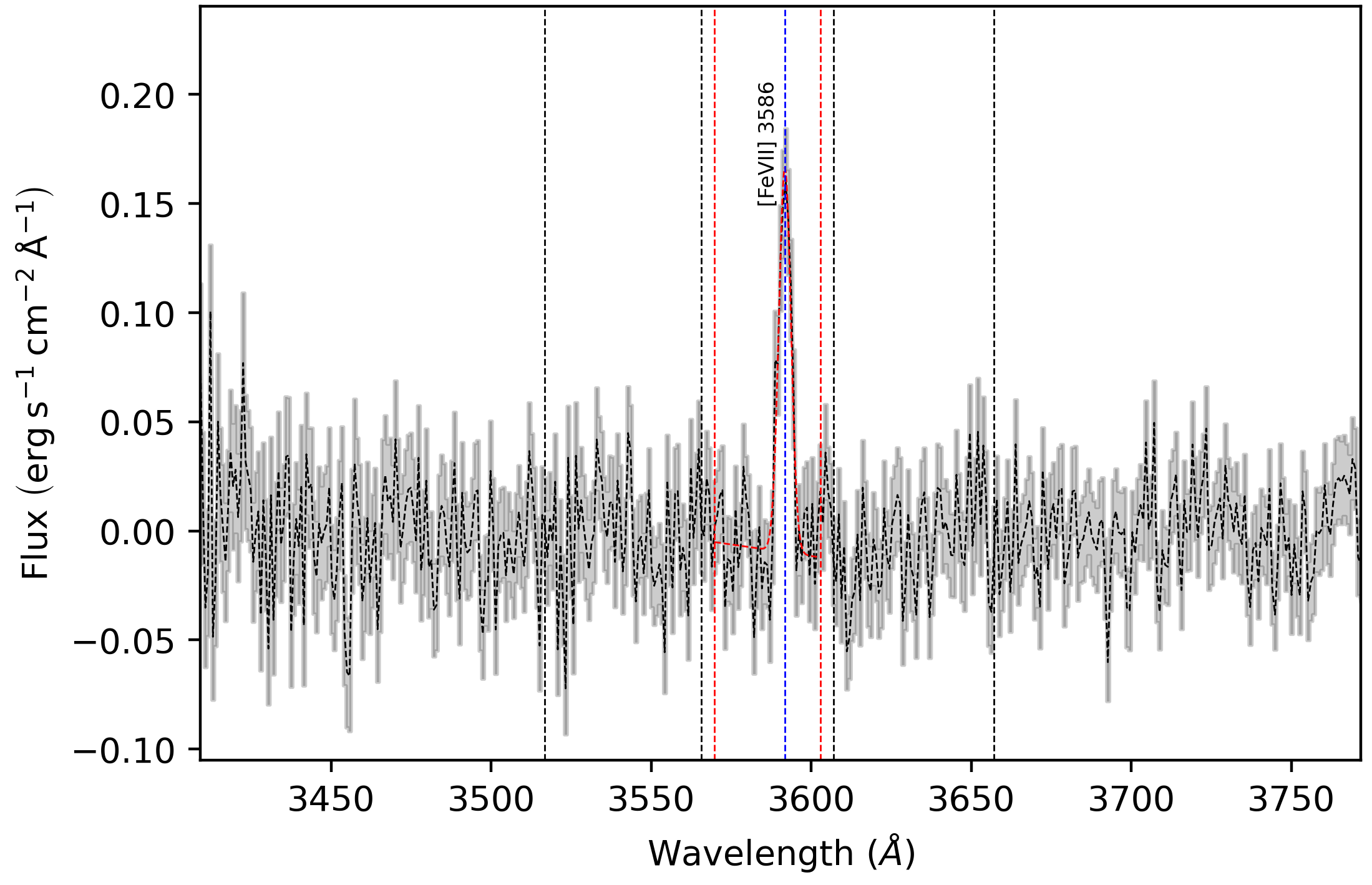}
	\caption{A sample spectrum from an individual spaxel showing the [FeVII]$\lambda3586$ line detected at $\ge 5 \sigma$ above the continuum in J0906. The dotted black line is the continuum subtracted spectrum, the shaded gray region is the uncertainty, the solid red line represents the best fit, the red dotted vertical lines mark the fitting window, the blue dotted line signifies the rest wavelength of the [FeVII]$\lambda3586$ line, and the two sets of black  dotted vertical lines correspond to the neighboring continuum windows where the RMS flux values of the continuum are calculated.}
	\label{fig:spec1}
\end{figure}

Once the spectra are stellar continuum subtracted, we attempt a single Gaussian ﬁt on a $\approx$ 30 \AA $\>$ region centered on the rest wavelengths of the CLs ([FeX]$\lambda$6374 being the exception; see Section \ref{fexfit}). We found that this wavelength range is adequate for capturing the full extent of CL emission in our preliminary scans. We then determine the root mean square (RMS) ﬂux of two continuum regions ($\approx$ 60 \AA $\>$wide) that neighbor each target CL, free of absorption or emission lines, and require that CL amplitudes are detected at $\ge$ 5$\sigma$ above the mean RMS ﬂux values in these continuum regions. We consider the spectral resolution of MaNGA (R = $\lambda$/$\Delta \lambda$ $\approx$ 1400 at 3600 \AA; R $\approx$ 2000 at 6000 \AA; \citealt{2013AJ....146...32S}) to eliminate ﬁts with $\Delta \lambda$ $\lessapprox$ 2.4 \AA $\>$(for [NeV]$\lambda$3347, $\lambda$3427), $\lessapprox$ 2.6 \AA (for [FeVII]$\lambda$3586, $\lambda$3760), and $\lessapprox$ 3 \AA (for [FeVII]$\lambda$6086 and [FeX]$\lambda$6374). We provide an example of a single Gaussian ﬁt for the [FeVII]$\lambda$3586 line in Figure \ref{fig:spec1}.
\subsubsection{[FeX] Emission Line Measurements}
\label{fexfit}
For [FeX]$\lambda6374$, the broad blue wing of this line is often blended with [OI]$\lambda6364$ due to their close proximity. Consequently, we attempt a double Gaussian fit to isolate the [FeX]$\lambda6374$ line. If this routine does not successfully fit both lines with $\ge$ 5$\sigma$ confidence, then we attempt a single Gaussian fit and apply the method used in \cite{2009MNRAS.397..172G} and \cite{2015MNRAS.448.2900R}, whereby the emission line ratio [OI]$\lambda6300$/$\lambda6364$ is used to determine if the [OI]$\lambda6364$ and [FeX]$\lambda6374$ lines are blended. Specifically, from atomic physics, if [OI]$\lambda6300$/$\lambda6364$ $= 3$, then the [OI]$\lambda6364$ line is free from contamination (see also \citealt{2011AcA....61..179E} for a full review). If [FeX]$\lambda6374$ emission is present and blended with [OI]$\lambda6364$, it will reduce the [OI]$\lambda6300$/$\lambda6364$ ratio below three. The MaNGA DAP provides flux values for both [OI]$\lambda$6364 and [OI]$\lambda$6300 lines. We adopt this method and require this ratio to be below three when fitting for [FeX]$\lambda6374$ with a single Gaussian fit to avoid confusing [OI]$\lambda6364$ and [FeX]$\lambda6374$ emission. Once we isolate the [FeX]$\lambda6374$ emission, we impose the same thresholds used to identify the [NeV] and [FeVII] emission lines (e.g., amplitudes $\ge$ 5$\sigma$; Section \ref{emline}).
\begin{table}[t]
	\renewcommand{\thetable}{\arabic{table}}
	\centering
	\caption{Target CLs} 
	\begin{tabular}{cccc}
		\hline
		\hline
		Emission Line &  Wavelength &IP& \textit{z}$_{\rm{min}}$  \\ {} & {(\AA)} & (eV) & {} \\
		\hline
		\rm{[NeV]} &    3347 &   126.2 &  0.088\\
		\rm{[NeV]} &    3427 &   126.2 & 0.061\\
		\rm{[FeVII]} & 3586     & 125.0   & 0.016\\
		\rm{[FeVII]} &  3760   &  125.0   & -\\
		\rm{[FeVII]} &  6086   &   125.0  & -\\
		\rm{[FeX]} &  6374   &  262.1  & -\\
		\hline
		\multicolumn{4}{p{8cm}}
		{Note: Columns are (1) emission line, (2) rest wavelength, (3) ionization potential, and (4) minimum redshift value required for MaNGA detection.}
	\end{tabular}
	\label{tab:coronal}
\end{table} 
\subsection {Coronal Line Flux Maps}
\label{clflux}
\begin{figure}[t]
	\includegraphics[height = 2.8in]{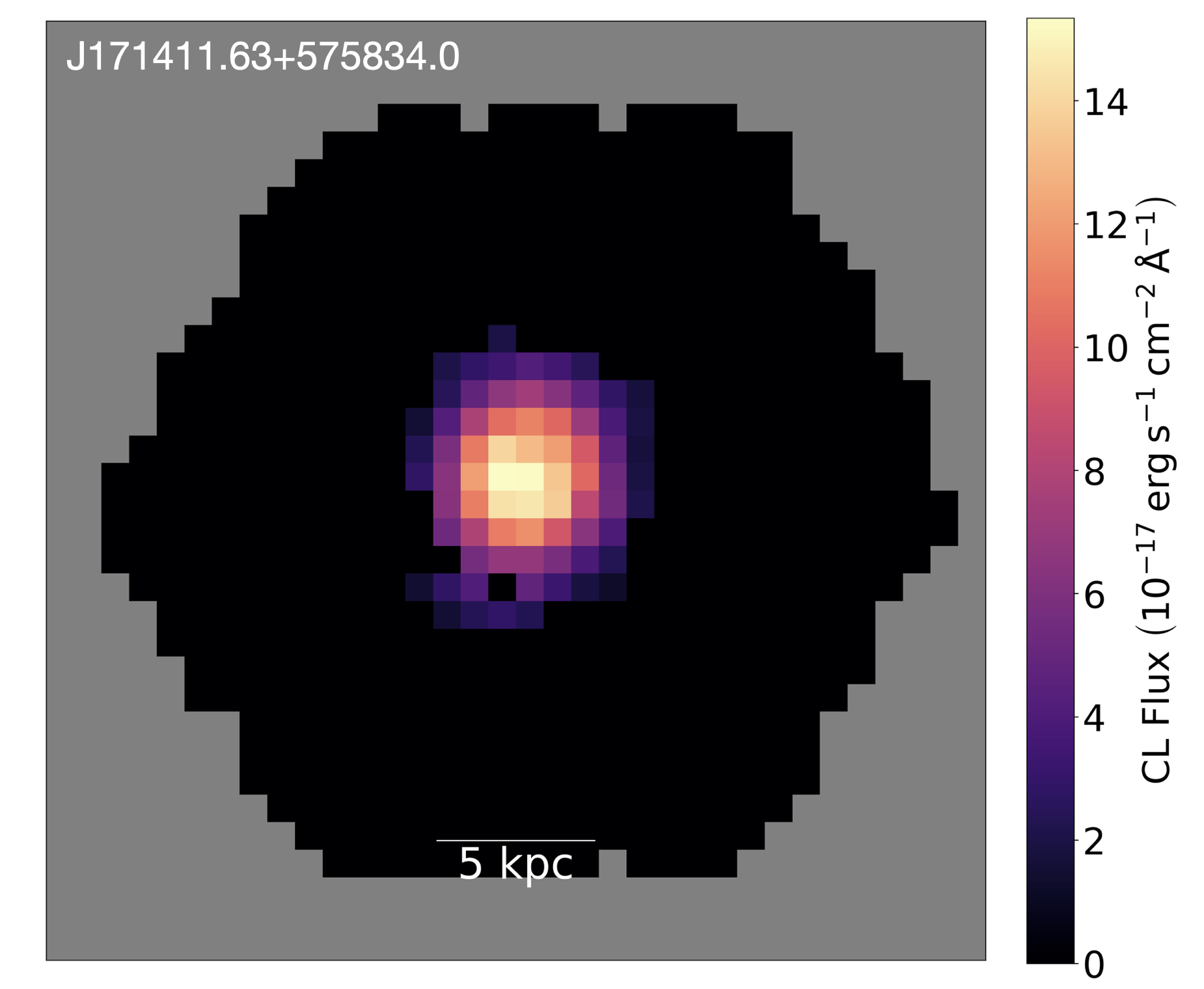}
	\caption{A sample CL flux map showing [NeV]$\lambda3427$ emission detected $\ge 5 \sigma$ above the continuum in J$1714$. For this galaxy, the strongest [NeV]$\lambda3427$ emission is located near the center of the galaxy. The gray region is outside of the MaNGA FoV and the black region are spaxels with no CL emission. North is up, south is down, east is to the left, and west is to the right.}
	\label{examplefluxmap}
\end{figure}

Similar to \citealt{2021ApJ...920...62N}, we create custom CL ﬂux maps to analyze the strength and distribution of the CLs in the CLR. We create these maps using the integrated CL ﬂux value from each spaxel for each CL galaxy (Figure \ref{examplefluxmap}). 

The center of each MaNGA observation corresponds to the galactic center (\citealt{2016AJ....152..197Y}). We use this position and the galaxy’s inclination angle to determine the de-projected galactocentric distance of each CL spaxel. We do acknowledge that the CL gas may not be restricted to the galactic disk; i.e., the CL emission may associated with an ionization ``cone" and therefore, in these instances, the de-projected distances are approximations. 

The MaNGA DAP provides the ratio of the semi-minor to semi-major axes (b/a) for each galaxy, and we use this value to determine the cosine of each galaxy’s inclination angle (i): cos(i)= b/a. The de-projected distance of each CL spaxel to the center of the galaxy is then measured by:
\begin{equation}
	\label{eqn1}
	\begin{aligned}
	\rm{CLD}= \sqrt{(x - x_{\rm{center}})^{2} + \big((y - y_{\rm{center}})*\rm{cos}(i) \big)^{2}}
	\end{aligned}
\end{equation}
where x is the projected distance between the spaxel and the galaxy center measured along the galaxy's major axis, and y for the minor axis.

We then convert spaxel distances to a physical unit (kpc) using the \texttt{astropy.cosmology} Python package. The resulting value corresponds to the coronal line distance (CLD) of each CL emitting spaxel from the galactic center. Further, the minimum coronal line distance (CLD$_{\rm{min}}$) corresponds to the distance of each galaxy’s closest CL-emitting spaxel from the galactic center. Finally, the maximum coronal line distance (CLD$_{\rm{max}}$) corresponds to the distance of each galaxy’s most distant CL-emitting spaxel from the galactic center.
\subsection {Galaxy Morphology}
\label{morph}
To uncover the correlation, if any, between CL emission and galaxy morphology (e.g., spiral and elliptical), we use the MaNGA Morphologies Galaxy Zoo value-added catalog to classify the morphologies of the galaxies in our sample. This catalog features data from Galaxy Zoo 2, a ``citizen science'' catalog with more than 16 million visual morphological classiﬁcations for $>$ 304,000 galaxies in SDSS (GZ2; \citealt{2013MNRAS.435.2835W}), 

The weighted vote fraction (discussed in \citealt{2013MNRAS.435.2835W}) accounts for voter consistency when participants select morphological classifications, and we require this fraction to be $\ge$ 50\% before assigning a morphological classiﬁcation (e.g., “E” for elliptical, or ``S" for spiral). We also use the weighted vote fraction to determine if a CL galaxy features a bar, and/or is categorized as odd (``b" and``o", respectively).

In addition, to determine the fraction of CL galaxies undergoing a merger, we consider the analysis being performed by Nevin et al., in prep (``Nevin catalog" hereafter). The authors determine the merger probability for each of the 1.3 million galaxies in the SDSS DR16 photometric sample, using a statistical learning tool that is built on a linear discriminant analysis framework, which is trained to separate mock images of simulated merging and non-merging galaxies using imaging predictors (see \citealt{2019ApJ...872...76N} for a full review). We investigate the MPL-11 galaxies from the broader SDSS DR16 Nevin catalog, and classify a CL galaxy as a merger if the Nevin catalog gives it a merger value (p$_{\rm{merg}}$) $>$ 0.5. 
\subsection {AGN Bolometric and [OIII] Luminosities}
\label{lum}
The AGN bolometric luminosity effectively traces the energetic output of an AGN (across the entire electromagnetic spectrum). To compare the luminosity of the CL AGN candidates with other known AGN candidates, we thus consider the bolometric luminosity parameter. 

We determine the AGN bolometric luminosity for each CL galaxy using the summed [OIII] ﬂux values ($F_{\rm{[OIII]}} $) across the entire galaxy (provided by the MaNGA DAP), and the procedure outlined in \cite{2017MNRAS.468.1433P}, which assumes [OIII] emission comes from an AGN:
\begin{equation}
	\label{eqn2}
	\begin{aligned}
		\rm{log}\>\left(\frac{L_{\rm{bol}}}{erg s^{-1}}\right) = (0.5617 \pm 0.0978)\>\rm{log}\> \left(\frac{L_{[OIII]}}{erg s^{-1}}\right)\\ + (21.186 \pm 4.164)
	\end{aligned}
\end{equation}
where $L_{\rm{[OIII]}} =  F_{\rm{[OIII]}} (4\pi R^{2})$ and R is the DAP provided luminosity distance based on redshift and a standard cosmology of $\Omega_{M}$ = 0.3 and $\Omega_{\Lambda}$ = 0.7 (redshift is also measured by the DAP). 

We then measure the total [OIII] luminosity (using the summed [OIII] fluxes across the entire galaxy) for each CL galaxy in our sample. Next, we compare the [OIII]  luminosities  of the CLs in our pipeline (Section \ref{bol_oiii}) to determine the relative strength of [OIII] for each CL. We do so to assess if specific CLs are preferentially found in higher or lower luminosity [OIII]-emitting galaxies, which is useful to determine if CLs uniformly trace all AGN, or if there may be an [OIII] luminosity dependence. 
\subsection{Narrow-Line BPT Diagnostics Diagrams}
\label{bptsec}

Baldwin-Phillips-Terlevich optical emission-line diagnostic diagrams (BPT diagrams; \citealt{1981PASP...93....5B,1987ApJS...63..295V, 2001ApJ...556..121K, 2006MNRAS.372..961K}) are widely accepted to be effective tools for categorizing gas ionization sources as star-forming, Seyfert (AGN), low-ionization nuclear emission-line region (LINER), or a composite of multiple ionization sources. They serve as the traditional AGN selection tool for most spectroscopic surveys. Specifically, these diagrams compare line ratios between high and low ionization species, most commonly [OIII]$\lambda$5007/H$\beta$ vs. [NII]$\lambda$6583/H$\alpha$ (``[NII]/ H$\alpha$ diagram" hereafter).

In this paper, we construct spatially resolved narrow-line BPT diagnostic diagrams for the CL galaxies to better constrain the ionization sources of the CLs. To do so, we require emission line measurements for the [NII]$\lambda$6583, [OIII]$\lambda$5007, H$\alpha$, and H$\beta$ emission lines. The DAP measures the continuum subtracted flux for each of these emission lines. Note, these fluxes account for galactic reddening using the E(B-V) values determined by the DRP, which assumes an \cite{1994ApJ...422..158O} reddening law. 

\begin{figure}[t]
	\includegraphics[height = 3.3in]{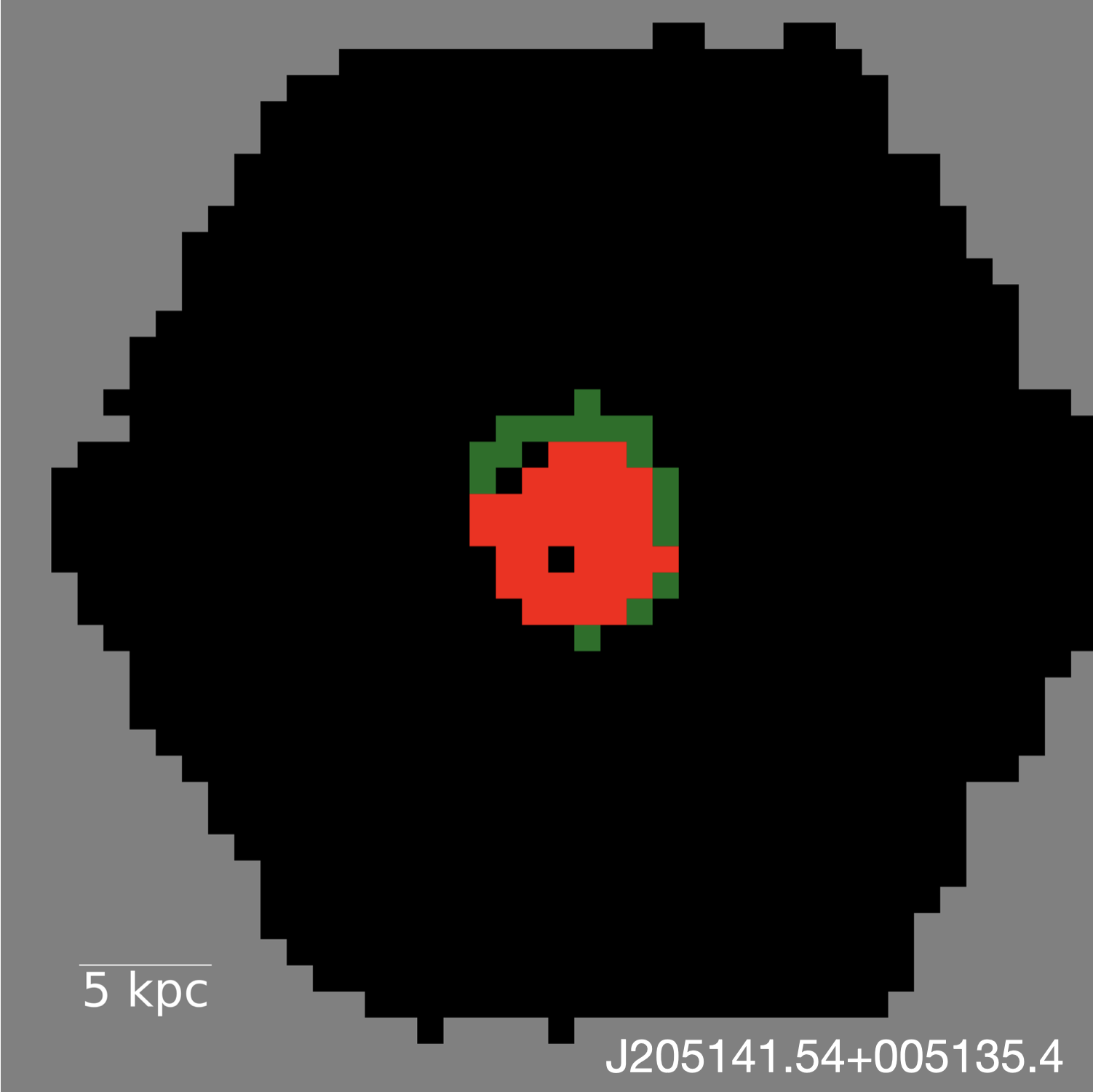}
	\caption{A sample BPT map showing AGN spaxels in red and composite spaxels in green, for CL-emitting spaxels in J$2051$ (a [NeV]$\lambda$3427 galaxy). The gray region is outside of the MaNGA FoV and the black region are spaxels with no CL emission. North is up, south is down, east is to the left, and west is to the right.}
	\label{bpt}
\end{figure}

Once we determine the necessary emission line flux measurements, we compute the ratios for the [NII]/ H$\alpha$ diagram, for each CL-emitting spaxel. We then use these values to create custom spatially-resolved BPT maps, whereby we present the BPT-classification for each CL-emitting spaxel within the MaNGA FoV, for each CL galaxy. Figure \ref{bpt} shows an example BPT map. 

\subsection{Dust Attenuation}
\label{dust}
\cite{2009MNRAS.394L..16M} investigated the [FeVII]$\lambda$6086, [FeX]$\lambda$6374, and [FeXI]$\lambda$7892 emission lines in the Seyfert 1 galaxy Ark 564. The authors used the photoionization code CLOUDY (\citealt{1998PASP..110..761F}) to determine the location and kinematics of these lines. They found that the CLs are launched from a dusty torus near the SMBH, where the gas is quickly accelerated. Moreover, using the CLOUDY models, they determined that some iron carrying grains are destroyed during the initial acceleration of the gas. 

To follow up on the analysis performed by \cite{2009MNRAS.394L..16M}, and to better understand the role of dust grains on the potential depletion of the iron CLs, we use the E(B - V) color excess index. This index traces the degree of interstellar reddening caused by photons that are scattered off of dust; in essence, it measures the difference between an object's observed color index and its intrinsic color index.  E(B - V) values for each CL galaxy are provided by the MaNGA DRP (using \citealt{1998ApJ...500..525S} maps), and assume the extinction law provided by
\cite{1994ApJ...422..158O}. 
\subsection{Shock Diagnostics}
\label{shockssec}
We explore the role of shocks (e.g., supernova remnant (SNR) and [OI]$\lambda$6300 (``[OI]" hereafter) shocks) in our analysis to elucidate the role of collisional excitation in the production CLs (e.g., \citealt{1984MNRAS.208..347P}). To do so, we consider the strength of the [SII]$\lambda$6717, $\lambda$6731 doublet with respect to the H$\alpha$ line, which has traditionally been used to differentiate SNR shocks from photoionized regions. Specifically,  \cite{1978MmSAI..49..485D} and \cite{1980A&AS...40...67D} first determined  that regions with [SII] ($\lambda6717  + \lambda6731$)/H$\alpha$ $>$ 0.4 can be used to identify SNR shocks. Additionally, the [OI] emission line is generally a strong tracer of shock excitation, and line ﬂux ratios with [OI]/H$\alpha$$>$ 0.1 indicate that shocks with velocities 160-300 km s $^{-1}$ are the main excitation source of [OI] (e.g., \citealt{1976ApJ...209..395D,2008ApJS..178...20A,2010ApJ...724..267F,2010ApJ...721..505R, 2011ApJ...734...87R, 2021MNRAS.501L..54R, 2022ApJ...927...23C}). The MaNGA DAP provides flux measurements for the [SII]$\lambda$6717, $\lambda$6731, [OI], and H$\alpha$ emission lines. 

\section{Results}
\label{sec:results}
In this section, we report the main findings for the CL galaxies in our sample. First, we present the fraction of confirmed AGN in the CL galaxies. Then, we analyze the spatial distribution and extent of the CLs. Next, we inspect the CL galaxy bolometric and [OIII] luminosities to deduce the effectiveness of using each species for accurate AGN identification. After, we assess the BPT classification of the CL-emitting spaxels. Finally, we investigate the role of dust extinction and shocks in the CLR to determine the impact of dust grains on CL emission, and to further constrain the ionization source(s) of the CLs.

In total, we find 71 galaxies with CL emission at $\ge$ 5$\sigma$ above the background continuum in MaNGA’s MPL-11 (33 feature [NeV] emission, 39 feature [FeVII] emission, and 4 feature [FeX] emission). Note, in our sample, 40 unique CL galaxies with either [NeV]$\lambda$3427, [FeVII], or [FeX] emission, or a combination of the three, feature redshifts below the \textit{z}$_{\rm{min}}$ threshold for [NeV]$\lambda$3347 (\textit{z}$_{\rm{min}} = 0.088$); further, 24 unique CL galaxies with either [FeVII] or [FeX] emission feature redshifts below the \textit{z}$_{\rm{min}}$ threshold for [NeV]$\lambda$3427 (\textit{z}$_{\rm{min}} =0.061$). Therefore, we are unable to scan for [NeV]$\lambda$3347 or  [NeV]$\lambda$3427 in these respective galaxies. In general, most of the MPL-11 galaxies feature redshifts that place [NeV]$\lambda3347$, $\lambda3427$ out of MaNGA's spectral range (see Section \ref{cont}).

Moreover, in light of the extensive work to detect AGNs using infrared CLs, we crossmatched our catalog of 71 unique CL galaxies with the infrared CL catalogs presented in \citealt{1998ApJ...498..579G}, \citealt{2002A&A...393..821S},
 		\citealt{2004ApJS..154..178A} \citealt{2005ApJ...632L..13L},
 		\citealt{2006ApJ...651..101W}, \citealt{2008ApJ...674L...9D},
 		\citealt{2009MNRAS.398.1165G},
 		 \citealt{2022arXiv220913125A}. We do not identify any of the MaNGA CL galaxies in these samples.

For 63/71 CL galaxies with GZ2 classifications  (89\%), we determine a nearly even fraction of spirals and ellipticals (48\% and 52\%, respectively). In addition, we measure the average size of the CLR (from the galactic center) for [NeV]$\lambda$3347, $\lambda$3427, [FeVII]$\lambda$3586, $\lambda$3760, $\lambda$6086, and [FeX]$\lambda$6374 to be 1.9 kpc, 2.3 kpc, 3.7 kpc, 5.3 kpc, 4.1 kpc, and 2.5 kpc, respectively (Table \ref{spatial_table}). Further, we find that the vast majority of [NeV] galaxies feature at least one CL-emitting spaxel in their nuclear regions (98.5\%; 2.$^{\prime\prime}$5 x 2.$^{\prime\prime}$5 FoV surrounding the central spaxel), whereas [FeVII] and [FeX] galaxies generally feature a smaller fraction (73\% and 75\%, respectively). The corresponding fraction of confirmed AGN in these galaxies (determined by comparing our sample to the largest catalog of confirmed MaNGA AGN; Section \ref{agn}) is 94\%, 14\%, and 25\%, respectively.

\subsection{MaNGA AGN Comparison}
\label{agn}
Comerford et al. (in prep) provide the most complete sample of AGN in MaNGA's MPL-11 (see \citealt{2020ApJ...901..159C} for a full review of their MPL-8 MaNGA AGN catalog). The authors compile a catalog of MaNGA AGN that were detected using SDSS broad emission lines, NVSS/ FIRST 1.4 GHz radio observations, \textit{WISE} mid-infrared color cuts, and \textit{Swift}/BAT hard X-ray observations. 


Broad Balmer emission lines (FWHM $>$ 1,000 km s$^{-1}$) are strong tracers of the rapidly rotating, high density gas, near the SMBH. They serve as reliable tracers for AGN activity. \cite{2015ApJS..219....1O} assembled a catalog of nearby ($z$ $\le$ 2) Type I AGN in SDSS’s seventh data release using the broad H$\alpha$ emission line, and Comerford et al. (in prep) identify 78 broad line AGN from this catalog in MPL-11.

Powerful AGN radio jets can expand several kpcs from the SMBH, and can thus serve as strong signatures for AGN activity. As a result, detecting the radio emission from these sources is a great tool for accurate AGN identification. \cite{2012MNRAS.421.1569B} used observations from the 1.4 GHz NRAO Very Large Array Sky Survey (NVSS; \citealt{1998AJ....115.1693C}) and the Faint Images of the Radio Sky at Twenty Centimeters (FIRST; \citealt{1995ApJ...450..559B}) to detect AGN in the SDSS’s seventh data release (DR7). They diﬀerentiated AGN activity from star formation emission using the correlation between the 4000 \AA $\>$ break strength and radio luminosity per stellar mass, emission line diagnostics, and the relation between H$\alpha$ and radio luminosity (\citealt{1995ApJ...450..559B}). Comerford et al. (in prep) find 221 radio AGN from this catalog in MPL-11.

Heated dust that surrounds an AGN can produce mid-infrared emission, which can expose obscured and unobscured AGN activity. Comerford et al. (in prep) thus rely upon observations from the \textit{Wide-Field Infrared Survey Explorer} (\textit{WISE}; \citealt{2010AJ....140.1868W}) to help identify AGN. They consider the four bands observed with \textit{WISE} (3.4 $\mu$m (W1), 4.6 $\mu$m (W2), 12 $\mu$m (W3), and 22 $\mu$m (W4)) and apply a 75\% reliability criteria of W1 - W2 $>$ 0.486 exp\{0.092(W2 - 13.07)$^{2}$\} and W2 $>$ 13.07, or W1 - W2 $>$ 0.486 and W2 $\le$ 13.07  (\citealt{2018ApJS..234...23A}) to select AGN. Comerford et al. (in prep) detect 130 \textit{WISE} AGN in MPL-11.

X-ray emission produced by AGN generally result from inverse Compton scattering of low energy UV photons by energetic electrons from the accretion disk (e.g., \citealt{1993ARA&A..31..473A, 1991ApJ...380L..51H, 2021MNRAS.506.4960H}). Therefore, X-rays can be a useful indicator of AGN activity. Accordingly, the authors use the X-ray catalog assembled by \cite{2018ApJS..235....4O}, which consists of $\approx$ 1,000 AGN observed by the \textit{Swift Observatory’s} Burst Alert Telescope (BAT) in the ultra hard X-ray (14 - 195 keV), to detect AGN. Comerford et al. (in prep) uncover 30 AGN from this catalog in MPL-11.

We compare our CL sample to the AGN catalog reported by Comerford et al. (in prep; ``Comerford sample" hereafter) and crossmatch 35 CL galaxies in it (52\% of our sample). Further, we consider the fraction of CL galaxies with confirmed AGN by specific CL species. We determine that 94\% (31/33) of the [NeV] galaxies host an AGN; 14\% (5/36) of the [FeVII] galaxies and 25\% (1/4) of the [FeX] galaxies. Overall, 35 unique CL galaxies host a confirmed AGN; 80\% (28/35) are confirmed with \textit{WISE} observations, 63\% (22/35) with broad Balmer emission lines, 14\% (5/35) with NVSS observations, and 11\% (4/35) with BAT AGN.

All of the [NeV]$\lambda$3347 galaxies feature an AGN and [NeV]$\lambda$3427 emission, and of the five [FeVII] galaxies with a confirmed AGN, two (J0736 and J1714) also feature both [NeV]$\lambda$3347,  $\lambda$3427 emission. Further, two of the remaining three [FeVII] galaxies with a confirmed AGN (J0807 and J1157) feature emission from more than one [FeVII] emission line (J0807 features [FeVII]$\lambda$3586, $\lambda$3760, $\lambda$6086 emission; J1157 features [FeVII]$\lambda$3586, $\lambda$3760 emission). The final [FeVII] galaxy with a confirmed AGN exclusively features [FeVII]$\lambda$6086 emission, and the sole [FeX] galaxy with a confirmed AGN (J1628) exclusively features [FeX] emission. 

Provided that 80\% of the CL galaxies in our sample are confirmed to host an AGN via \textit{WISE} diagnostics, we consider the fact that these \textit{WISE} diagnostics are likely to miss low luminosity AGN (e.g., \citealt{2018ApJS..234...23A}). Perhaps, one possible explanation for the discrepancy in AGN across the CLs in our sample is that [NeV] traces high-luminosity AGN, while [FeVII] and [FeX] may possibly trace low-luminosity AGN. We explore this  further in Sections \ref{bol_oiii} and \ref{dust_1}.
\subsection{Spatial Distribution and Extent of the CLs}
\label{spat}
\begin{table*}
	\renewcommand{\thetable}{\arabic{table}}
	\centering
	\caption{Spatial Properties for the CL Galaxies} 
	\begin{tabular}{ccccccc}
		\hline
		\hline
		Detected&  Wavelength & Confirmed  & Nuclear Emission & {CLD$_{\rm{min}}$} & {CLD$_{\rm{max}}$} & {CLD$_{\rm{avg}}$}  \\ {CL} & {(\AA)} & {Galaxies} & (\%) & {kpc} & {kpc} & {kpc} \\
		{(1)}&{(2)}&{(3)}&{(4)}&{(5)}&{(6)}&{(7)}\\
		\hline
		\rm{[NeV]} &    3347 & 8 &  100& 0.52&  5.3 &  1.9\\
		\rm{[NeV]} &    3427 & 33 &   97 & 0.34  &  19 & 2.3\\
		\rm{[FeVII]} & 3586   & 4 & 100 & 0.10 & 9.6& 3.7 \\
		\rm{[FeVII]} &  3760   & 16 & 56& 0.11 & 36& 5.3 \\
		\rm{[FeVII]} &  6086   &19 &   63&0.10  &   21& 4.1\\
		\rm{[FeX]} &  6374   & 4 & 75 &0.60 & 4.9&  2.5\\
		\hline
		\multicolumn{7}{p{13cm}}
		{Note: Columns are (1) detected CL, (2) rest wavelength, (3) number of galaxies with CL emission detected, (4) percentage of  CL galaxies with at least one CL-emitting spaxel in a nuclear  2.$^{\prime\prime}$5 FoV, (5) the average CLD   (distance of CL-emitting spaxel from the galaxy center), (6) the distance of the furthest CL-emitting spaxel from the galaxy center, and (7) the distance of the closest CL-emitting spaxel from the galaxy center.}
	\end{tabular}
	\label{spatial_table}
\end{table*}
To better constrain the ionization source(s) of the CLs, we first map the measured fluxes of the CLs within the MaNGA FoV for each CL galaxy (Figure \ref{examplefluxmap}). These flux maps provide a snapshot of the orientation, extent, and intensity of CL emission for the galaxies in our sample. Then, we compute the de-projected distance of each CL spaxel from the nuclear center of each galaxy (i.e. the photometric center; Section \ref{clflux}) to determine the distance of each CL-emitting spaxel from the galaxy center. Finally, we define the nuclear region of each CL galaxy to be a 2.$^{\prime\prime}$5 x 2.$^{\prime\prime}$5 aperture (5 x 5 spaxel grid; where each spaxel covers a 0.$^{\prime\prime}$5 x  0.$^{\prime\prime}$5 FoV) surrounding the central spaxel. 

If AGN photoionization is the primary mechanism producing the CLs, it is likely that CL emission is predominantly within the nuclear region of each galaxy, close to the SMBH and the accretion disk (e.g., \citealt{2009MNRAS.397..172G,2010MNRAS.405.1315M}). On the other hand, if shocks, AGN outflows, or stellar processes play an active role in generating CLs, we anticipate that CL emission will not be found exclusively in the nuclear region. Rather, we would expect to find emission in regions off-center or off-axis from the SMBH and the galaxy's rotational plane (see \citealt{2021ApJ...920...62N} for more discussion). 

To analyze the CL distribution within the nuclear region of the CL galaxies, we measure the fraction of CL galaxies with at least one CL emitting spaxel in their center, for each CL (Table 	\ref{spatial_table}).  We find that the vast majority of [NeV] $ \lambda$3347, $\lambda$3427 galaxies feature at least one  [NeV]-emitting spaxel in their nuclear regions (100\% and 97\%, respectively). This finding is consistent with our results in Section \ref{agn}, that [NeV] is a strong tracer of AGN activity (i.e. CL emission is likely dominated by AGN photoionization near the SMBH). Comparatively, the fraction of CL galaxies with nuclear emission from [FeVII] $ \lambda$3586, $\lambda$3760, or $\lambda$6086 varies significantly more (100\%, 56\%, and 63\%, respectively). 

\begin{figure*}
	\includegraphics[height = 4.8in]{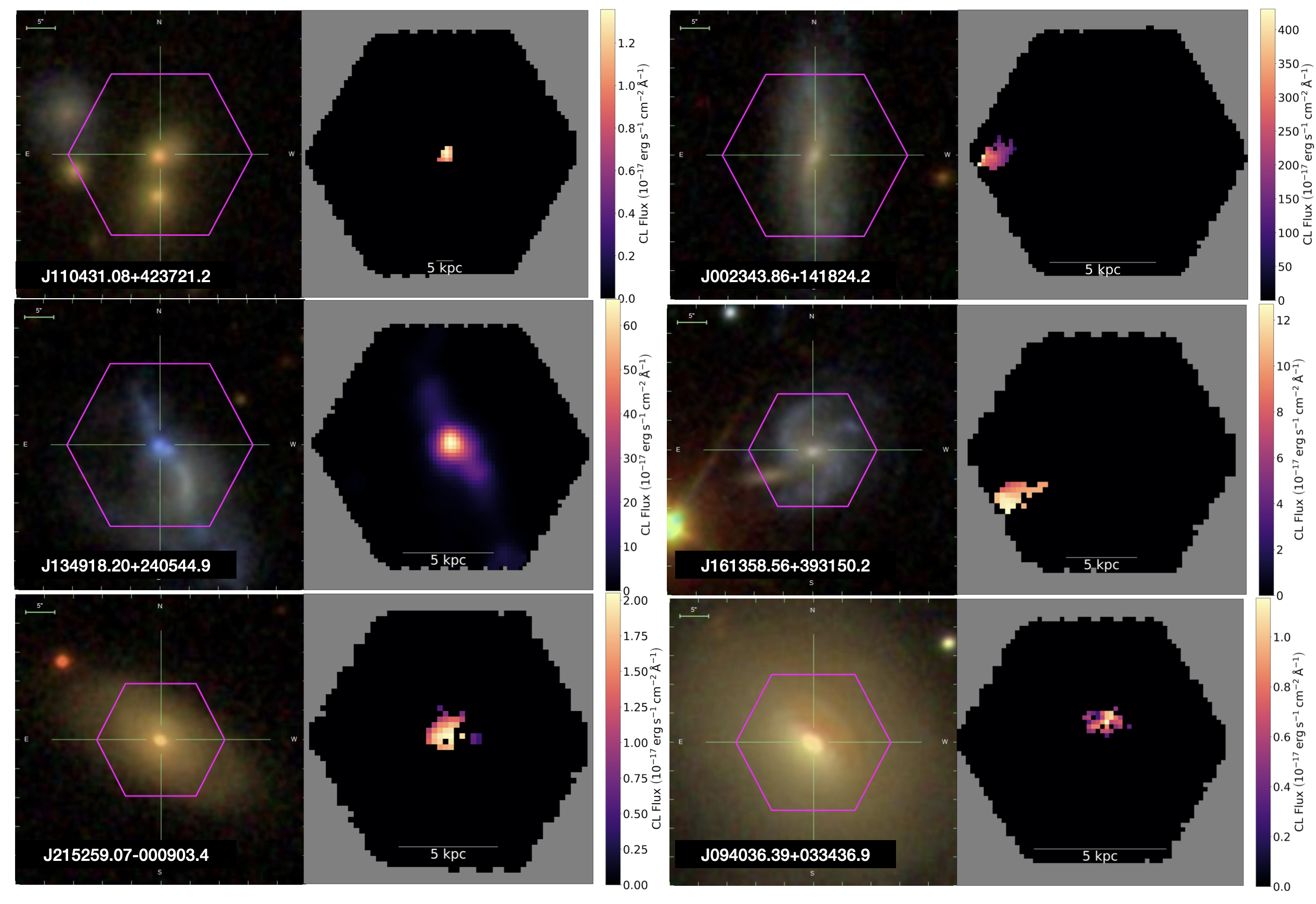}
	\caption{CL flux maps for 6/71 CL galaxies in our sample. From top to bottom and left to right: J1104 ([NeV]$\lambda$3427 map), J0023 ([FeVII]$\lambda$3760 map), J1349 ([FeVII]$\lambda$3760 map), J1613 ([FeVII]$\lambda$3760 map), J2152 ([FeVII]$\lambda$6086 map), and J0940 ([FeVII]$\lambda$6086 map). For J0023 and J0940, the maps display CL emission spatially offset from the galaxy center. For each galaxy, the emission is offset perpendicular to the rotational plane of the galaxy, suggestive of the source of the CLs being AGN outflows. For J1349 we observe a possible companion galaxy and consider the possibility that these two galaxies to be undergoing a merger. For the J1613 observation, we determine that several optical emission lines (e.g., [OIII] and H$\alpha$) measured in the secondary galaxy (with the featured ``CL emission"; southwest of J1613 in the MaNGA FoV) have large velocity shifts ($>$ 2,000 km s$^{-1}$) compared to the center of J1613, which suggests this may not be a companion galaxy (i.e., this is not a merging system; J1613 is not in the Nevin catalog). As a result, the ``CL emission" in this galaxy is likely from a separate emission line, from a background galaxy with a different redshift than the primary galaxy. For  J1104  and J2152, CL emission is concentrated towards the galaxy center, likely produced by AGN photoionization.}
	\label{fig:spatial}
\end{figure*}

In Figure \ref{fig:spatial}, we present a sample of CL flux maps for six representative CL galaxies. In three of the galaxies (J1104, J1349, and J2152), it is apparent that the source of the CLs is within the nuclear region, as the CL flux is concentrated here. However, for the three remaining galaxies (J0023, J1613, and J0920), the CL-emitting spaxels are highly offset from the nuclear region. Based on the orientation of the CL flux in J0023 and J0920, it is possible that AGN outflows are generating the CL since the CL emission is generally perpendicular to the orbital plane of each galaxy. For the J1613 observation, we determine that several optical emission lines (e.g., [OIII] and H$\alpha$) measured in the secondary galaxy (with the featured ``CL emission"; southwest of J1613 in the MaNGA FoV) have large velocity shifts ($>$ 2,000 km s$^{-1}$) compared to the center of J1613, which suggests this may not be a companion galaxy (i.e., this is not a merging system; J1613 is not in the Nevin catalog). As a result, the ``CL emission" in this galaxy is likely from a separate emission line, from a background galaxy with a different redshift than the primary galaxy. For J1349, we acknowledge that there appears to be a visual companion galaxy near the nuclear region; it is possible that both merger induced shocks and AGN photoionization could be producing the CL emission. In addition, we suspect that dust grains may also have a significant impact on the presence of [FeVII] and [FeX] emission in our sample. In Section \ref{dust_1}, we review the likelihood of iron depletion by dust grains more thoroughly.

We also compute the CLDs for the CL galaxies. The CLD, which is the distance of each CL-emitting spaxel from the galactic center, reveals the physical scale of the CLR for each CL galaxy. We measure the average CLDs for [NeV]$\lambda$3347,  $\lambda$3427 to be 1.9 kpc and 2.3 kpc, respectively; 3.7 kpc, 5.3 kpc, and 4.1 kpc for [FeVII]$\lambda$3586, $\lambda$3760, $\lambda$6086, respectively; 2.5 kpc for [FeX]$\lambda$6374. We find no correlation between IPs and CLDs (IP = 262.1 eV for [FeX], IP  = 126.2 eV for [NeV], and IP = 125 eV for [FeVII]). Moreover, for the [NeV] galaxies, the minimum and maximum distances of each CL-emitting spaxel from the nuclear center (labeled CLD$_{min}$ and CLD$_{max}$ in Table \ref{spatial_table}) ranges between 340 pc to 19 kpc, 100 pc to 36 kpc for the [FeVII] galaxies, and 600 pc to 4.9 kpc for the [FeX] galaxies. These large variances in CL distance suggest that the CLR extends from just beyond the BLR ($\approx$ 0.1 kpc) and well into the NLR (several kpcs). 

Finally, to confirm that CL emission is indeed resolved for each CL galaxy, we consider the instrument PSF ($\approx$ 2.$^{\prime\prime}$5 for MaNGA), and find that 60/71 CL galaxies show resolved and continuous emission in excess of the typical instrument PSF. The remaining 11 CL galaxies (J0205, J1010, J1117, J1317, J1344, J1416, J1604, J1626, J1628, J1658, and J1649) lack CL emission in excess of the typical instrument PSF. We reason that these CLRs are below the instrument PSF, and not spatially resolved. As discussed in \cite{2021ApJ...920...62N}, these CLRs may still be spatially resolved by other instruments (e.g., \citealt{2010MNRAS.405.1315M} and their use of STIS/HST optical spectra), and it is also posible that CL emission may be oriented along an ionization cone; however here we consider the CLDs of these galaxies to be upper limits.
\subsection{AGN Bolometric and [OIII] Luminosities}
\label{bol_oiii}
AGN bolometric luminosity, which scales with [OIII] luminosity, is effectively the ``power" of an AGN. As outlined in \cite{2017MNRAS.468.1433P}, [OIII] emission is the most utilized line for measuring bolometric luminosity, due its strength in most AGN spectra and the relatively weak blending of emission from photoionized gas in star forming regions with the line (e.g., \citealt{2004ApJ...613..109H, 2014ARA&A..52..589H}).

\begin{figure}[t]
	\includegraphics[height = 3in]{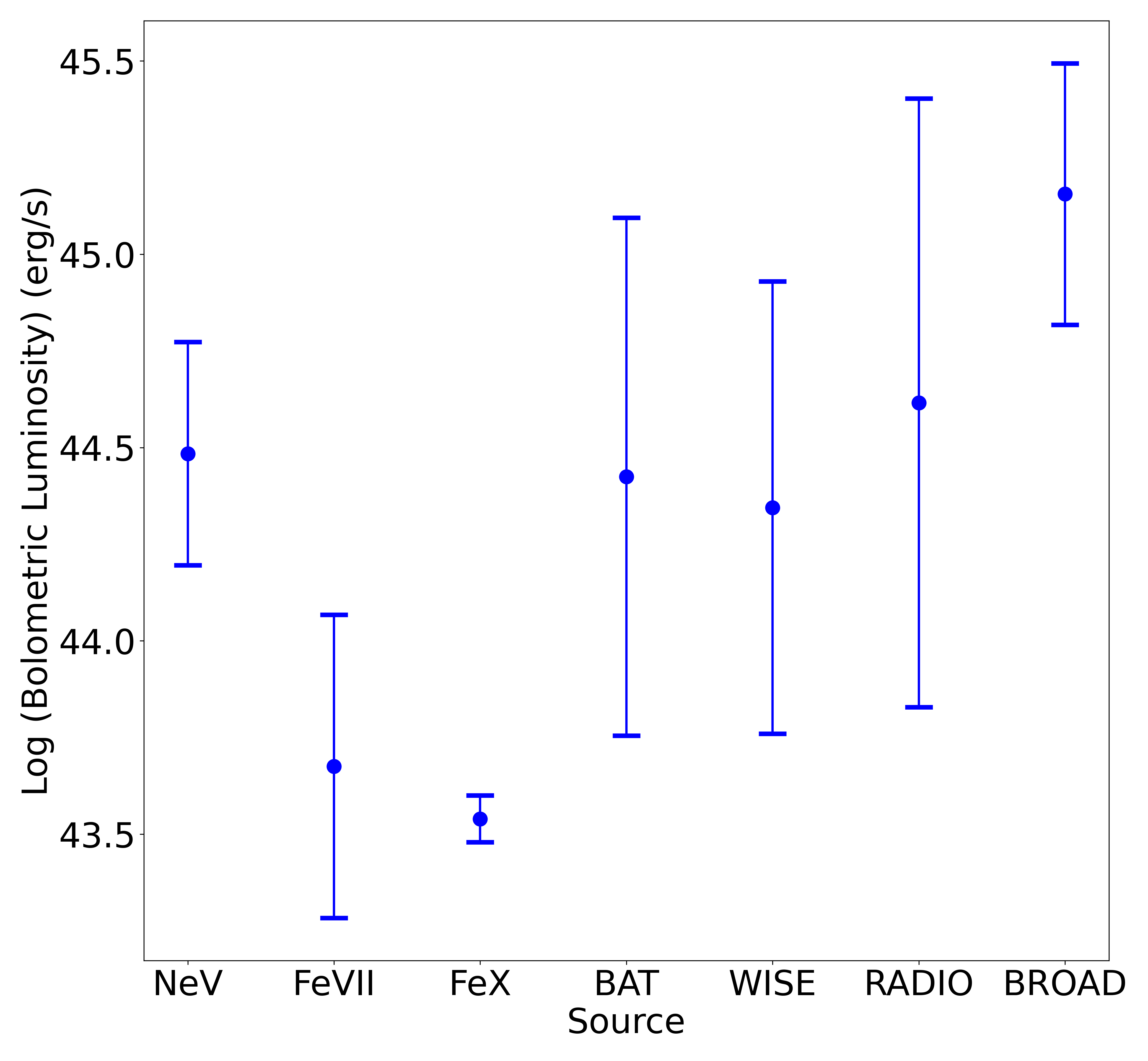}
	\caption{Average mean bolometric luminosities for the 71 CL galaxies in our sample (analyzed by each CL species; [NeV], [FeVII], and [FeX]), compared to the MaNGA galaxies confirmed to feature an AGN in the Comerford sample. The AGN in the Comerford sample were verified using SDSS broad emission lines, NVSS/ FIRST 1.4 GHz radio observations, \textit{WISE} mid-infrared color cuts, and \textit{Swift}/BAT hard X-ray observations. The mean bolometric luminosity of the [NeV] galaxies, log($L_{\rm{bol}}$) $= 44.5$ erg s$^{-1}$, is consistent with the AGN reported in the Comerford sample (mean log($L_{\rm{bol}}$) = 44.6 erg s$^{-1}$ for the Comerford sample). However, the [FeVII] and [FeX] galaxies feature mean bolometric luminosities an order of magnitude lower ($\rm{log}(L_{\rm{bol}}) \le 43.7$ erg s$^{-1}$) than the Comerford sample. We suspect that the [FeVII] and [FeX] emission lines may primarily be detecting low luminosity AGN in MaNGA.}
	\label{bol_mean}
\end{figure}

Therefore, to help resolve the discrepancy between the differing fractions of confirmed AGN in our sample  (Section \ref{agn}; 94\% of the [NeV] galaxies feature a confirmed AGN, 14\% for the [FeVII] galaxies, and 25\% of the [FeX] galaxies; for CL galaxies with multiple CLs, we measure this fraction independently for each CL), and to evaluate the overall effectiveness of using CL detections to identify AGN, we consider the bolometric and [OIII] luminosities of the CL galaxies, and further inspect the Comerford sample of MaNGA AGN. In particular, we compare the mean bolometric luminosities of the CL galaxies ($L_{\rm{bol}}$; using the summed [OIII] flux across the entire galaxy; Section \ref{lum}) with the total population of MPL-11 AGN in the Comerford sample (Section \ref{agn}; Figure \ref{bol_mean}). 

\begin{figure}[t]
	\includegraphics[height = 3in]{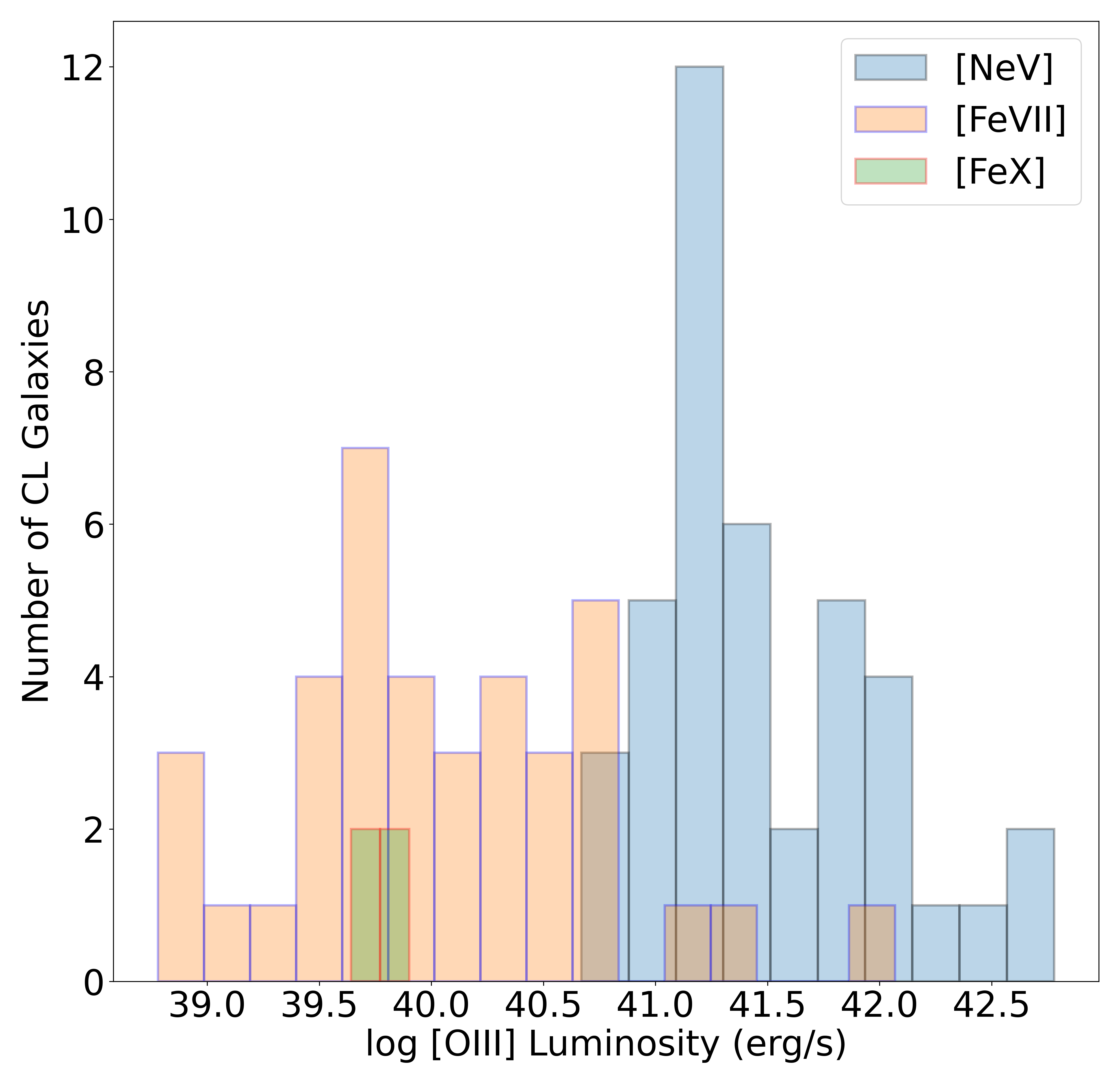}
	\caption{The log([OIII]) luminosity distribution for the 71 CL galaxies. The blue, orange, and green histograms represent the [NeV], [FeVII], and [FeX] galaxies in our sample, respectively. [NeV] emitting galaxies tend to have higher [OIII] luminosities than [FeVII] or [FeX], which suggests that these galaxies may host higher luminosity AGN. The mean of the [NeV] log([OIII]) luminosity distribution is 41.5 erg s$^{-1}$; 40.1 erg s$^{-1}$ and 39.8 erg s$^{-1}$ for [FeVII] and [FeX], respectively.}
	\label{fig:oiii}
\end{figure}

We find that the mean bolometric luminosity for the [NeV] galaxies (mean \textit{z} = 0.10; median \textit{z} = 0.11), log($L_{\rm{bol}}$) $= 44.5$ erg s$^{-1}$, is consistent with the mean value of Comerford's population of MaNGA galaxies that host an AGN (log($L_{\rm{bol}}$) = 44.6 erg s$^{-1}$). On the other hand, we measure the mean bolometric luminosities of the [FeVII] galaxies (mean \textit{z} = 0.06; median \textit{z} = 0.05) and the [FeX] galaxies (mean \textit{z} = 0.07; median \textit{z} = 0.06) to be an order of magnitude lower than the mean log($L_{\rm{bol}}$) value of the Comerford sample (log($L_{\rm{bol}})$ $=$ 43.7 erg s$^{-1}$ and log($L_{\rm{bol}}$) $=$ 43.5 erg s$^{-1}$ for the [FeVII] and [FeX] galaxies, respectively). Note, we also present the [OIII] luminosity distribution for the CL galaxies in Figure \ref{fig:oiii} (the mean [OIII] luminosity for the [NeV]  galaxies is 41.5 erg s$^{-1}$; 40.1 erg s$^{-1}$ and 39.8 erg s$^{-1}$ for the [FeVII] and [FeX] galaxies, respectively). We reason that the [FeVII] and [FeX] galaxies may be preferentially tracing lower luminosity AGN in MaNGA, which are generally more difficult to detect in multi-wavelength observations. 

However, we find that the five [FeVII] galaxies with a confirmed AGN (J0736, J0807, J1157, J1535, and J1714) all feature relatively high [OIII] luminosities of log(L$_{\rm{[OIII]}}$)  $\gtrapprox$ 41 erg s$^{-1}$. Additionally, the three remaining [FeVII] galaxies with [OIII] luminosities at or above this limit (without confirmed AGN) are J0906, J1349, and J2152. Both J0906 and J1349 visually appear to be actively undergoing a merger; J2152 shows no apparent companion galaxy. We reason that, for the [FeVII] galaxies in our sample, the log(L$_{\rm{[OIII]}}$) cutoff of $\approx$ 41 erg s$^{-1}$ is a useful threshold for identifying confirmed AGN (from the Comerford sample) and may also be helpful for detecting mergers. Further, for the [NeV] galaxies, J1344 features the lowest [OIII] luminosity (log(L$_{\rm{[OIII]}}$)  $=$ 40.6 erg s$^{-1}$) and in fact hosts a confirmed AGN. We consider the [OIII] luminosity threshold for the [FeVII] galaxies (log(L$_{\rm{[OIII]}}$) $\approx$ 41 erg s$^{-1}$) to be similar for the [NeV] galaxies. 

We also determine that the two [NeV] galaxies (J1658 and J1104) that do not feature a confirmed AGN (out of 33 total [NeV] galaxies), feature [OIII] luminosities of log(L$_{\rm{[OIII]}}$) $=$ 41.2 erg s$^{-1}$ and log(L$_{\rm{[OIII]}}$)  $=$ 41.4 erg s$^{-1}$, respectively. Considering these high [OIII] luminosities, and the high [NeV] AGN detection rate (94\%), we propose that these two galaxies are strong AGN candidates.

On the other hand, the one [FeX] galaxy with a confirmed AGN, J1628, features an [OIII] luminosity of log(L$_{\rm{[OIII]}}$) $=$ 39.8 erg s$^{-1}$ (the remaining three [FeX] galaxies, which do not host a confirmed AGN, also have log([OIII]) luminosities $<$ 40 erg s$^{-1}$). Consequently, while we consider the log(L$_{\rm{[OIII]}}$) $\approx$ 41 erg s$^{-1}$ threshold useful for identifying CL galaxies with a confirmed AGN, it is important to acknowledge that CL galaxies with a confirmed AGN can have [OIII] luminosities below this limit.  

\subsection{BPT Analysis} 
 \begin{table*}
		\renewcommand{\thetable}{\arabic{table}}
		\centering
		\caption{BPT Classifications for the CL Galaxies}. 

		\begin{tabular}{ccccccccc}
			\hline
			\hline
			Detected  &  Rest & Confirmed  & Confirmed & Confirmed  & [NII] &[NII]  & [NII]  \\
			{CL} & {Wavelength} &{CL Galaxies} & {AGN} & {AGN Fraction} &{AGN Fraction} & {Composite Fraction} &{SF Fraction}  \\
			{} & {\AA} & {} & {} & {\%} & {\%} & {\%} & {\%}  \\
			{(1)} & {(2)} & {(3)} & {(4)} &  {(5)} & {(6)} &  {(7)} & {(8)}   \\
			\hline
			
			[NeV]  & $\lambda$3347 &8& 8 & 100& 87.5 &12.5 &0\\
			{}& $\lambda$3427 & 33 & 31 & 94& 90&10 & 0\\
			{[FeVII]}  & $\lambda$3586 & 4 & 3 & 75 & 78.5 & 19 &2.5 \\
			{}& $\lambda$3760 & 16&2 &13& 80.3 & 5.8 {}&13.9 \\
			&  $\lambda$6086  &19& 3 & 16 & 67.9&19.7&12.5\\
			{[FeX]}& $\lambda$6374  &4& 1&25 & 88.3 &0 &11.7\\
			\hline 
			\multicolumn{9}{p{17cm}}
			{Note: Columns are (1) detected CL, (2) rest wavelength, (3) the number of  galaxies that feature emission from the respective line, (4) the fraction of galaxies that host a confirmed AGN, (5) the average fraction of [NII] AGN BPT spaxels, (6) the average fraction of [NII] Composite BPT spaxels, and (7) the average fraction of [NII] SF BPT spaxels.}
		\end{tabular}
		\label{tab:bpt}
	\end{table*} 
The BPT diagram has long served as the standard tool for identifying ionization mechanisms in emission line sources (e.g, \citealt{1981PASP...93....5B,1987ApJS...63..295V, 2001ApJ...556..121K, 2006MNRAS.372..961K}). While effects such as stellar shocks and emission from post-AGB stars are liable to elevate SF sources beyond the AGN threshold (see \citealt{2012ApJ...747...61Y, 2016MNRAS.461.3111B, 2019ApJ...876...12A} for a further discussion), we nonetheless explore the BPT classification for each CL-emitting spaxel to help pin down the source of CL emission in our sample. To do so, we compute the log([OIII]/H$\beta$) and log([NII]/H$\alpha$) ratios required for the [NII]/H$\alpha$ diagram (Section \ref{bptsec}). Using the thresholds outlined in \cite{2006MNRAS.372..961K}, we categorize each CL spaxel as either [HII] (i.e. star forming), AGN, or a composite of the two. Note, for some CL-spaxels, the DAP reports negative values for the necessary emission line fluxes, likely because the emission lines of interest yield low flux levels and the DAP’s subtraction of the stellar continuum results in a net absorption at the expected wavelength of the emission line. As such, we exclude these spaxels from our analysis. 

For the [NeV], [FeVII], and [FeX] emission lines, we determine that, on average, the majority of CL-emitting spaxels are AGN or composite (Table \ref{tab:bpt}). For [NeV]$\lambda$3347, $\lambda$3427 we find that 87.5\% and 90\% of these spaxels are classified as AGN, respectively; 12.5\% and 10\% composite, respectively; 0\% SF for both. Moreover, we measure the BPT ratios for the [FeVII]$\lambda$3586, $\lambda$3760, $\lambda$6086 spaxels, and find that 78.5\%, 80.3\%, and 67.9\% of these spaxels are classified as AGN, respectively; 19\%, 5.8\%, and 19.7\% composite, respectively; 2.5\%, 13.9\%, and 12.5\% SF, respectively. For [FeX]$\lambda$6374, 88.3\% of the CL-emitting spaxels are classified as AGN; 0\% composite;  11.7\% SF. In total, 100\% of the [NeV] spaxels in our sample are either BPT AGN or BPT composite,  91\% of the [FeVII] spaxels, and 88.3\% of the [FeX] spaxels. These results suggest that the CLs are perhaps useful tracers of AGN, and that the lack of confirmed AGN in our [FeVII] and [FeX] galaxies may trace back to the nearly bimodal log([OIII]) and bolometric luminosity distributions presented in Section \ref{bol_oiii} (i.e. [FeVII] and [FeX] may generally be found in low luminosity AGN that are potentially missed by traditional AGN detection techniques; though, it is also possible that [FeVII] and [FeX] may not host an AGN at all). We explore this possibility, and the corresponding impact of dust extinction on iron CL emission in Section \ref{dust_1}.

 \subsection{The Impact of Dust on CL Emission}
\label{dust_1}
The role of dust extinction (i.e. the  impact of dust grains) on CL emission has yet to be fully unraveled. \cite{2009MNRAS.394L..16M} suggest that dust grains can potentially deplete heavier CL species (e.g., iron). Further, \cite{1997ApJS..110..287F}  posit that there are three primary effects of dust on line formation: 1) emission lines weaken due to the absorption of the incident continuum by dust, 2) grains photoelectrially heat the gas, and 3) some of the gas-phase elements (e.g., iron) are depleted (see also \citealt{1983ApJ...275..652S, 1996ApJ...468L..65S, 2009ApJ...694..765C, 2009ApJ...698..106K}). Comparatively,  \cite{1997ApJS..110..287F} contend that neon (a noble gas; i.e. a species of gas with a full outer shell of valence electrons, and thus less chemical reactivity) is significantly less depleted by dust grains, and therefore [NeV] is emitted almost fully outside the grain sublimation radii. Here, we consider the likelihood that a significant population of iron CL photons are destroyed by dust in our sample. 

To explore the role of dust extinction on CL emission in our sample, and to determine its relevance for the discrepancy between the fraction of confirmed AGN in the [NeV] galaxies (94\%) vs. the [FeVII] and [FeX] galaxies (14\% and 25\% respectively; Section \ref{agn}), we use the E(B - V) color excess index, which traces interstellar reddening (Section  \ref{dust}). The  MaNGA DRP provides this index for each galaxy in MPL-11, and we use it to determine if there is a correlation between the dust content of each CL galaxy and its CL emission. We present our findings in Table \ref{dust_table}.

In particular, we find the mean E(B - V) values for the [FeVII]$\lambda$3760, $\lambda$6086 galaxies to be the lowest (i.e. feature less dust grains) across our sample (0.029 and 0.039, respectively). Because iron is susceptible to destruction by dust grains, particularly in the nuclear region where the presence of dust is greater (also due to dust in the NLR), these relatively  low values provide a viable explanation for the presence of [FeVII]$\lambda$3760, $\lambda$6086 emission in these galaxies; for reference, our sample contains 16 [FeVII]$\lambda$3760 galaxies and 19 [FeVII]$\lambda$6086 galaxies. Comparatively, the [NeV]$\lambda$3347, $\lambda$3427, [FeVII]$\lambda$3586, and [FeX]$\lambda$6374 galaxies feature higher mean E(B- V) values; 0.057, 0.045, 0.045, and 0.049, respectively. The corresponding number of iron CL-emitting galaxies found in these galaxies is only nine in total (J0736 features emission from both [NeV] lines, as well as [FeVII]$\lambda$3586 emission; J1714 features emission from both [NeV] lines, as well as [FeVII]$\lambda$6086 emission; J0807 features emission from [FeVII]$\lambda$3586, $\lambda$3760, $\lambda$6086; J1157 features emission from [FeVII]$\lambda$3586, $\lambda$3760; J0906 features emission from [FeVII]$\lambda$3586; J1628, J2311, J1649, and J1720 exclusively feature emission from [FeX]$\lambda$6374). We suspect that emission from the iron CL species is being diminished within these relatively dusty galaxies, which provides a physical explanation for the low number of iron CL galaxies in the high E(B - V) value galaxies (E(B - V) $\ge$ 0.045; nine iron CL galaxies) vs. the low E(B - V) galaxies (E(B - V) $<$ 0.039; 34 iron CL galaxies).
\begin{table}[t]
	\renewcommand{\thetable}{\arabic{table}}
	\centering
	\caption{CLR Dust Attenuation} 
	\begin{tabular}{cccc}
		\hline
		\hline
		Detected&  Wavelength & Average E(B-V) Value  \\ {CL} & {(\AA)} & {} & {}  \\
		{(1)}&{(2)}&{(3)}\\
		\hline
		\rm{[NeV]} &    3347 & 0.057 \\
		{} &    3427 & 0.045\\
		\rm{[FeVII]} & 3586   & 0.045 \\
		{} &  3760   & 0.029\\
		{} &  6086   &0.039 \\
		{\rm{[FeX]}} &  6374   & 0.049  \\
		\hline
		\multicolumn{3}{p{6cm}}
		{Note: Columns are (1) detected CL, (2) rest wavelength, and (3) average E(B-V) values, for each CL, reported by MaNGA's DRP.}
	\end{tabular}
	\label{dust_table}
\end{table}  

Furthermore, \citealt{2006ApJ...648L.101E} considered the correlation between the AGN dusty torus and AGN bolometric luminosity. They proposed that the dusty torus diminishes at log($L_{\rm{bol}})$ $\lessapprox$ 42 erg s$^{-1}$, due to mass accretion no longer being able to sustain the necessary cloud outflow rate, which effectively results in a decrease in column density (see also \citealt{1999A&A...349...77C, 2004ApJ...602..116W, 2005ApJ...625..699M}). While the cloud component of the AGN  is not immediately extinguished below this threshold, the authors contend that the cloud outflow rate at  log($L_{\rm{bol}})$ $\lessapprox$ 42 erg s$^{-1}$ is less than the necessary ``standard" observed in higher luminosity AGN. As a result, we consider the L$_{\rm{bol}}$ values for the CL species (Figure	\ref{bol_mean}, mean $\rm{log}(L_{\rm{bol}}) \ge 44.3$ erg s$^{-1}$ for [NeV] galaxies; mean $\rm{log}(L_{\rm{bol}}) \le 43.7$ erg s$^{-1}$ for the  [FeVII] and [FeX] galaxies) to conclude that the lower L$_{\rm{bol}}$ values correlate with a diminishing dusty torus, which results in less destruction of iron by dust grains. Accordingly, we detect more iron CLs in these low luminosity sources. On the other hand, the [NeV] galaxies feature higher L$_{\rm{bol}}$ values, which likely correspond to their elevated E(B - V) values. Likewise, since L$_{\rm{bol}}$ scales with L$_{\rm{OIII}}$, this reasoning elucidates the nearly bimodal log(L$_{\rm{OIII}}$) distribution of the [NeV] vs. the [FeVII] and [FeX] galaxies (Figure \ref{fig:oiii}; the mean of the [NeV] log([OIII]) luminosity distribution is 41.5 erg s$^{-1}$; 40.1 erg s$^{-1}$ and 39.8 erg s$^{-1}$ for [FeVII] and [FeX], respectively). 

  \subsection{SNR, [OI], and Merger-Induced Shocks in the CLR}
  \label{shcks}
   \begin{table*}
  	\renewcommand{\thetable}{\arabic{table}}
  	\centering
  	\caption{Morphological and Merger Classifications of the CL Galaxies} 
  	\begin{tabular}{ccccc}
  		\hline
  		\hline
  		SDSS Name&  Detected CL(s) & Redshift & Morphology& Merger  \\ {} & {} & {} & {} & {}  \\
  		{(1)}&{(2)}&{(3)}&{(4)} &{(5)}\\
  		\hline
J001938.78+144201.1 & [FeVII]$\lambda$6086 & 0.116 & E & N \\J002343.86+141824.2 & [FeVII]$\lambda$3760 & 0.018 & S(b) & N \\J020557.03+004623.9 & [FeVII]$\lambda$6086 & 0.042 & E & N \\J021257.59+140610.2 & [NeV]$\lambda$3427 & 0.062 & - & N \\J030639.57+000343.1 & [NeV]$\lambda$3427 & 0.107 & E  & Y \\J072656.07+410136.0 & [NeV]$\lambda$3427 & 0.129 & S(b) & Y \\J073623.13+392617.7 & [NeV]$\lambda$3347, [NeV]$\lambda$3427, [FeVII]$\lambda$3586 & 0.118 & - & Y \\J074128.48+442431.6 & [NeV]$\lambda$3427 & 0.132 & E & N \\J075217.84+193542.2 & [NeV]$\lambda$3427 & 0.117 & - & - \\J075756.71+395936.1 & [NeV]$\lambda$3427 & 0.066 & E(o) & Y \\J080018.53+461112.3 & [FeVII]$\lambda$3760 & 0.061 & E & N \\J080403.40+404809.3 & [NeV]$\lambda$3427 & 0.126 & S  & Y \\J080543.32+252710.9 & [FeVII]$\lambda$3760 & 0.072 & E & Y \\J080707.18+361400.5 & [FeVII]$\lambda$3586, [FeVII]$\lambda$3760, [FeVII]$\lambda$6086 & 0.032 & S & N \\J080859.19+364112.9 & [FeVII]$\lambda$6086 & 0.03 & E & Y \\J084002.36+294902.6 & [NeV]$\lambda$3427 & 0.065 & E & N \\J085208.48+511845.8 & [FeVII]$\lambda$3760 & 0.115 & S & N \\J085601.94+572327.4 & [FeVII]$\lambda$6086 & 0.041 & S(bo) & N \\J085835.98+013149.5 & [NeV]$\lambda$3427 & 0.107 & S(b) & Y \\J090659.46+204810.0 & [FeVII]$\lambda$3586 & 0.109 & S(o) & N \\J092002.85+054407.7 & [FeVII]$\lambda$6086 & 0.038 & S(b) & - \\J092739.77+050312.5 & [NeV]$\lambda$3427 & 0.126 & S & N \\J094036.39+033436.9 & [FeVII]$\lambda$6086 & 0.016 & - & Y \\J101042.59+061157.0 & [NeV]$\lambda$3427 & 0.098 & E(o) & Y \\J103825.16-002331.1 & [NeV]$\lambda$3347, [NeV]$\lambda$3427 & 0.096 & S(o) & Y \\J105439.31+475144.2 & [NeV]$\lambda$3427 & 0.073 & S(b) & N \\J105759.31+404940.6 & [FeVII]$\lambda$6086 & 0.024 & E & Y \\J110431.08+423721.2 & [NeV]$\lambda$3427 & 0.126 & E(o) & Y \\J111403.52+472653.4 & [FeVII]$\lambda$3760 & 0.113 & E & Y \\J111711.79+465134.0 & [FeVII]$\lambda$3760 & 0.061 & E & N \\J111724.94+443347.8 & [FeVII]$\lambda$3760 & 0.066 & E & N \\J111803.22+450646.8 & [NeV]$\lambda$3347, [NeV]$\lambda$3427 & 0.107 & E(o) & Y \\J112043.79+534337.4 & [FeVII]$\lambda$6086 & 0.107 & E & N \\J115710.68+221746.2 & [FeVII]$\lambda$3586, [FeVII]$\lambda$3760 & 0.052 & S(b) & N \\J122443.43+442438.8 & [NeV]$\lambda$3427 & 0.126 & E & Y \\J123521.03+422002.6 & [FeVII]$\lambda$6086 & 0.039 & E & N \\J130626.65+451720.4 & [FeVII]$\lambda$6086 & 0.051 & E & - \\J131730.11+474659.3 & [FeVII]$\lambda$3760 & 0.027 & E & N \\J134401.90+255628.3 & [NeV]$\lambda$3427 & 0.062 & S(b) & N \\J134918.20+240544.9 & [FeVII]$\lambda$3760 & 0.021 & - & Y \\J141623.14+381127.4 & [NeV]$\lambda$3427 & 0.135 & - & Y \\J142004.29+470716.8 & [NeV]$\lambda$3427 & 0.07 & S(b) & N \\J144454.24+522648.5 & [FeVII]$\lambda$3760 & 0.146 & E & N \\
  		
  		\hline
  		\multicolumn{5}{p{15cm}}
  		{Note: Columns are (1) SDSS Name, (2) detected CL(s), (3) redshift, (4) GZ2 morphological classifications; ``E" is for elliptical, ``S" is for spiral, ``b" is for bar, ``o'' is for odd, and ``-" indicates no morphological classification was assigned, and (5) the merger classification from the Nevin catalog; ``Y" marks galaxies with p$_{\rm{merg}}$ $>$ 0.5, ``N" identifies galaxies with p$_{\rm{merg}}$ $\le$ 0.5, and ``-" represents galaxies that are not in the Nevin catalog.}
  	\end{tabular}
  	\label{cl_galaxies}
  \end{table*}
  
  \begin{table*}
  	\renewcommand{\thetable}{\arabic{table}}
  	\centering
  	\caption{Morphological and Merger Classifications of the CL Galaxies (Continued)} 
  	\begin{tabular}{ccccc}
  		\hline
  		\hline
  		SDSS Name&  CL & Redshift & Morphology& Merger  \\ {} & {} & {} & {} & {} \\
  		{(1)}&{(2)}&{(3)}&{(4)} &{(5)}\\
  		\hline
  		J145420.10+470022.3 & [FeVII]$\lambda$3760 & 0.126 & E(o) & Y \\J151600.58+342119.1 & [NeV]$\lambda$3427 & 0.125 & S(o) & Y \\J151856.39+332152.2 & [FeVII]$\lambda$6086 & 0.069 & E & N \\J153552.40+575409.4 & [FeVII]$\lambda$6086 & 0.03 & E(o) & N \\J160455.20+280956.9 & [NeV]$\lambda$3427 & 0.077 & S & Y \\J161301.62+371714.9 & [NeV]$\lambda$3427 & 0.069 & S(b) & Y \\J161358.56+393150.2 & [FeVII]$\lambda$3760 & 0.038 & S & - \\J161413.20+260416.3 & [NeV]$\lambda$3347, [NeV]$\lambda$3427 & 0.131 & -(o) & Y \\J162428.39+483548.0 & [FeVII]$\lambda$6086 & 0.057 & E(o) & Y \\J162621.91+405442.7 & [FeVII]$\lambda$3760 & 0.03 & S(o) & Y \\J162845.89+252938.0 & [FeX]$\lambda$6374 & 0.04 & E & N \\J162908.95+383256.6 & [FeVII]$\lambda$6086 & 0.033 & E & N \\J163014.63+261223.3 & [NeV]$\lambda$3347, [NeV]$\lambda$3427 & 0.131 & S(b) & N \\J163053.84+243343.5 & [FeVII]$\lambda$6086 & 0.063 & E & Y \\J163430.87+374143.6 & [NeV]$\lambda$3427 & 0.099 & E & N \\J164956.39+351243.5 & [FeX]$\lambda$6374 & 0.1 & E & N \\J165810.10+622456.3 & [NeV]$\lambda$3427 & 0.119 & S(b) & N \\J171411.63+575834.0 & [NeV]$\lambda$3347, [NeV]$\lambda$3427, [FeVII]$\lambda$6086 & 0.093 & E & Y \\J172032.02+280602.9 & [FeX]$\lambda$6374 & 0.083 & - & Y \\J205141.54+005135.4 & [NeV]$\lambda$3347, [NeV]$\lambda$3427 & 0.106 & S  & Y \\J211646.34+110237.4 & [NeV]$\lambda$3427 & 0.081 & S & Y \\J212401.89-002158.6 & [NeV]$\lambda$3427 & 0.062 & S(b) & Y \\J212900.75+001057.3 & [FeVII]$\lambda$3760 & 0.133 & E(o) & N \\J213227.90+100816.9 & [NeV]$\lambda$3427 & 0.063 & S(o) & - \\J215259.07-000903.4 & [FeVII]$\lambda$6086 & 0.028 & S & N \\J223338.41+131243.6 & [NeV]$\lambda$3347, [NeV]$\lambda$3427 & 0.093 & S & Y \\J231142.05+150638.2 & [FeX]$\lambda$6374 & 0.04 & S(o) & N \\J232538.54+152115.8 & [FeVII]$\lambda$6086 & 0.041 & S & N \\
  		\hline
  		\multicolumn{5}{p{15cm}}
  		{Note: Columns are (1) SDSS Name, (2) detected CL(s), (3) redshift, (4) GZ2 morphological classifications; ``E" is for elliptical, ``S" is for spiral, ``b" is for bar, ``o'' is for odd, and ``-" indicates no morphological classification was assigned, and (5) the merger classification from the Nevin catalog; ``Y" marks galaxies with p$_{\rm{merg}}$ $>$ 0.5, ``N" identifies galaxies with p$_{\rm{merg}}$ $\le$ 0.5, and ``-" represents galaxies that are not in the Nevin catalog.}
  	\end{tabular}
  	\label{cl_galaxies_2}
  \end{table*}
  \begin{table*}
  	\renewcommand{\thetable}{\arabic{table}}
  	\centering
  	\caption{SNR Shocks} 
  	\begin{tabular}{cccccccccc}
  		\hline
  		\hline
  		Detected&  Wavelength & CL AGN & CL Non-AGN & CL Mergers & CL Non-Mergers    & CL Nuclear  & CL Non-Nuclear  \\ {CL} & {(\AA)} &  SNR Shocks & SNR Shocks &SNR Shocks & SNR Shocks & SNR Shocks & SNR Shocks \\
  		{(1)}&{(2)}&{(3)}&{(4)}&{(5)}&{(6)}&{(7)} & {(8)}\\
  		\hline
  		\rm{[NeV]} &    3347 & 13\% &  -& 14\%&  0\% &  13\% & -\\
  		\rm{[NeV]} &    3427 & 33\% &   100\% & 30\%  &  50\% & 39\% & 0\%\\
  		\rm{[FeVII]} & 3586   & 3\% & 38\% & 0\% &8\%& 12\% & - \\
  		\rm{[FeVII]} &  3760   & 6\% & 55\%& 49\% & 49\%& 52\% & 45\%\\
  		\rm{[FeVII]} &  6086   &95\% &   42\%&64\%  &   41\%& 61\% & 25\%\\
  		\rm{[FeX]} &  6374   & 100\% & 45\% &36\% & 67\%&  67\% & 36\%\\
  		\hline
  		\multicolumn{8}{p{17cm}}
  			{Note: Columns are (1) detected CL, (2) rest wavelength, (3) percentage of CL-emitting spaxels in the CL Galaxies with a confirmed AGN that feature SNR shocks, (4) percentage of CL-emitting spaxels in the CL Galaxies without a confirmed AGN that feature SNR shocks, (5) percentage of CL-emitting spaxels in the CL Galaxies undergoing a merger that feature SNR shocks, (6) percentage of CL-emitting spaxels in the CL Galaxies not undergoing a merger that feature SNR shocks, (7) percentage of CL-emitting spaxels in the CL Galaxies with nuclear CL emission (See Section \ref{spat})  that feature SNR shocks, and (8) percentage of CL-emitting spaxels in the CL Galaxies without nuclear CL emission  that feature SNR shocks. ``-" indicates an empty sample set.}
  	\end{tabular}
  	\label{shock_1}
  \end{table*}
  \begin{table*}
	\renewcommand{\thetable}{\arabic{table}}
	\centering
	\caption{[OI] Shocks} 
	\begin{tabular}{cccccccccc}
		\hline
		\hline
		Detected&  Wavelength & CL AGN & CL Non-AGN & CL Mergers & CL Non-Mergers    & CL Nuclear  & CL Non-Nuclear  \\ {CL} & {(\AA)} &  SNR Shocks & [OI] Shocks &[OI] Shocks & [OI] Shocks & [OI] Shocks & [OI] Shocks \\
		{(1)}&{(2)}&{(3)}&{(4)}&{(5)}&{(6)}&{(7)}\\
		\hline
		\rm{[NeV]} &    3347 & 25\% &  -& 29\%&  0\% &  25\% & -\\
		\rm{[NeV]} &    3427 & 29\% &   100\% & 33\%  &  31\% & 35\% & 0\%\\
		\rm{[FeVII]} & 3586   &4\% & 12\% & 0\% &6\%& 6\% & - \\
		\rm{[FeVII]} &  3760   & 3\% & 28\%& 20\% & 26\%& 25\% & 25\%\\
		\rm{[FeVII]} &  6086   &50\% &   36\%&45\%  &   33\%& 31\% & 49\%\\
		\rm{[FeX]} &  6374   & 100\% & 67\% &100\% & 67\%&  67\% & 100\%\\
		\hline
  		\multicolumn{8}{p{17cm}}
{Note: Columns are (1) detected CL, (2) rest wavelength, (3) percentage of CL-emitting spaxels in the CL Galaxies with a confirmed AGN that feature [OI] shocks, (4) percentage of CL-emitting spaxels in the CL Galaxies without a confirmed AGN that feature [OI] shocks, (5) percentage of CL-emitting spaxels in the CL Galaxies undergoing a merger that feature [OI] shocks, (6) percentage of CL-emitting spaxels in the CL Galaxies not undergoing a merger that feature [OI] shocks, (7) percentage of CL-emitting spaxels in the CL Galaxies with nuclear CL emission (See Section \ref{spat})  that feature [OI] shocks, and (8) percentage of CL-emitting spaxels in the CL Galaxies without nuclear CL emission  that feature [OI] shocks. ``-" indicates an empty sample set.}
	\end{tabular}
	\label{shock_2}
\end{table*}
 Astrophysical shocks can result from a variety of mechanisms, which include, but are not limited to, galaxy collisions, SNRs, cloud-cloud collisions, expanding HII regions, and outflows from young stellar objects (see \citealt{2008ApJS..178...20A} for a further review). To deduce the role of shocks in the CLR, we consider the [SII] ($\lambda6717  + \lambda6731$)/H$\alpha$ and  [OI]$\lambda$6300/H$\alpha$ ratios for each CL-emitting spaxel in our sample (values $>$ 0.4 indicate SNR shocks and values $>$ 0.1 trace [OI] shocks, respectively; Section \ref{shockssec}). 

We also investigate the fraction of CL galaxies actively undergoing a merger using the Nevin et al., catalog (Table \ref{cl_galaxies}; Table \ref{cl_galaxies_2}; Section \ref{morph}). In \citealt{2021ApJ...920...62N}, we found that the 3/10 CL galaxies without a confirmed AGN were all strong merger candidates (J0906, J1349, and J1454). Therefore, we consider the possibility that companion galaxies can drive gas inflows towards the galactic centers, resulting in merger-induced shock excitation (e.g., \citealt{2010ApJ...724..267F}) that may also produce CLs. Using the Nevin catalog, here we determine that 32/66 of the CL galaxies (48\%; 5 CL galaxies are not reported in the Nevin catalog: J0752, J0920, J1306, J1613, and J2132) have p$_{\rm{merg}}$ values $>$ 0.5 - indicative of an ongoing merger. 

Further, we present our SNR and [OI] shocks results in Tables \ref{shock_1} and \ref{shock_2}.  Overall, we find that the fraction of SNR and [OI] shocks do not vary significantly for the CL galaxies. In particular, on average and across all CL species, 42\% of the CL-emitting spaxels in the CL galaxies with a confirmed AGN feature SNR shocks (35\% feature [OI] shocks), whereas 56\% of the CL-emitting spaxels in the CL galaxies without a confirmed AGN feature SNR shocks (49\% feature [OI] shocks). Further, on average and across all CL species, 32\% of the CL-emitting spaxels in the CL galaxies undergoing a merger feature SNR shocks (38\% feature [OI] shocks). On the other hand, 36\% of the CL-emitting spaxels in the CL galaxies not undergoing a merger feature SNR shocks (27\% feature [OI] shocks). Finally, on average and across all CL species, 41\% of the CL-emitting spaxels in the CL galaxies with nuclear CL emission feature SNR shocks (32\% feature [OI] shocks). Comparatively, 27\% of the CL-emitting spaxels in the CL galaxies without nuclear CL emission feature SNR shocks (44\% feature [OI] shocks). 


We find clear evidence of SNR and [OI] shocks in the CL-emitting spaxels of each CL species in our sample. However, the fraction of these shocks does not strongly trace CL galaxies with or without: a confirmed AGN, a companion galaxy, or nuclear CL emission. We reason that  SNR and [OI] shocks may be viable CL-emission mechanisms; however, they are not likely dominant, and we find little evidence that they produce CLs  away from the nuclear region, in the absence of a confirmed AGN, or when a merging companion galaxy is present.

\section{Discussion}
\label{sec:discussion}
Based on our findings, we reason that the efficacy of using CLs to detect AGN varies by species of CL. While the ionization potential of each CL is $\ge$ 100 eV (Table \ref{tab:coronal}; well above the 55 eV threshold for pure star formation; consistent with the strong continuum of an AGN being the ionization source), we find that certain CLs are better at identifying higher luminosity AGN than others (log(L$_{\rm{[OIII]}}$)  $\gtrapprox$ 41 erg s$^{-1}$). In particular, [NeV] emission is predominately present in higher [OIII] luminosity galaxies that feature a confirmed AGN (mean log($L_{\rm{[OIII]}}$) $= 41.5$ erg s$^{-1}$ for the [NeV] galaxies; 94\% of the [NeV] galaxies host a confirmed AGN). On the other hand, we detect [FeVII] and [FeX] emission in lower [OIII] luminosity galaxies with fewer confirmed AGNs (mean $\rm{log}(L_{\rm{[OIII]}})$ $\le 40.1$ erg s$^{-1}$ for both CLs; 14\% and 25\%  confirmed AGN for the [FeVII] and [FeX] galaxies, respectively).

We reason that the destruction of iron CLs by dust grains, which we find is inversely proportional to AGN bolometric luminosity (the dusty torus diminishes at log($L_{\rm{bol}})$ $\lessapprox$ 42 erg s$^{-1}$; e.g., \citealt{2006ApJ...648L.101E}), may be directly impacting [FeVII] and [FeX] emission; the CL galaxies with the lowest E(B-V) values yield the most iron CL detections (nine iron CL galaxies with E(B - V) $\ge$ 0.045; 34 iron CL galaxies with (E(B - V) $<$ 0.039). We posit that if the [FeVII] and [FeX] galaxies host AGNs, that they may be lower luminosity AGNs, which are potentially too weak to be detected via SDSS broad emission lines, NVSS/ FIRST 1.4 GHz radio observations, \textit{WISE} mid-infrared color cuts, and \textit{Swift}/BAT hard X-ray observations.

We determine that there are primarily two distinct populations of CL galaxies in our sample: 1) a subset of CL galaxies that emit [NeV] (33/71 CL galaxies), with relatively high [OIII] and bolometric luminosities, and a high fraction of confirmed AGN (94\%), and 2) a group of CL galaxies that emit [FeVII] and [FeX] (40/71 CL galaxies), with relatively low [OIII] and bolometric luminosities, and a low fraction of confirmed AGN (14\% and 25\%, respectively).

Overall, we consider the similar IPs of [NeV] and [FeVII] (126.2 eV and 125 eV, respectively), the high IP of [FeX] (262.21 eV), and our BPT analysis (100\% of the [NeV] spaxels in our sample are either BPT AGN or NPT composite, 91\% of the [FeVII] spaxels, and 88.3\% of the [FeX] spaxels; Table \ref{tab:bpt}), to deduce that each CL in our sample is likely linked to AGN activity, but that [FeVII] and [FeX] emission may preferentially be found in less luminous AGN (we also consider the possibility that some of the iron CL emission may not exclusively be produced by AGN; e.g., shocks may also play a role). We conclude that the BPT diagram is generally effective at tracing large populations of AGN; however, [NeV], in particular, can also be used as an additional resource to help trace AGN, specifically for instances where one or more of the optical BPT line ratios is unable to be determined.  

Moreover, \cite{1997ApJS..110..287F} reported CL critical densities between 10$^{7}$ - 10$^{10}$ cm$^{-3}$, which suggest that the CLR is a region between the classical NLR and the BLR. The authors also indicate that lower ionization CLs (e.g., [NeV] and [FeVII]; IPs $\approx$ 125 eV) are more likely to form in lower density gas that should be spatially resolved. In contrast, higher ionization CLs (e.g., [FeX]; IP = 262.1 eV) form in a region closer to the nucleus where the ionizing flux, and ionization parameters, are higher (i.e., these CLs form in denser, more efficiently emitting regions). Here we determine that the average size of the CLR for [NeV]$\lambda$3347, $\lambda$3427, [FeVII]$\lambda$3586, $\lambda$3760, $\lambda$6086, and [FeX]$\lambda$6374 is 1.9 kpc, 2.3 kpc, 3.7 kpc, 5.3 kpc, 4.1 kpc, and 2.5 kpc, respectively (Table \ref{spatial_table}) - well into the NLR for all CL species. With the enhanced capabilities of IFS, which enables us to spatially resolve the CLR, we find that the CLR for the galaxies in MaNGA is larger than reported in previous works (tens to hundreds of pcs; e.g. \citealt{2005MNRAS.364L..28P, 2010MNRAS.405.1315M,  2011ApJ...739...69M, 2011ApJ...743..100R}).

Finally, while the bulk of the CL galaxies in our sample feature CL emission in their nuclear regions (within a central 2.$^{\prime\prime}$5 x 2.$^{\prime\prime}$5 FoV; Section \ref{spat}; Table \ref{spatial_table}), a significant fraction do not, which is inconsistent with pure AGN photoionization. In Section \ref{shcks}, we show that [OI] and SNR shocks are present in the CL-emitting spaxels of each CL species. However, the fraction of SNR and [OI] shocks, across all CL species, do not significantly vary for the CL galaxies with or without nuclear CL emission. We reason, instead, that AGN radio jets or outflows may be interacting with gas clouds away from the nuclear region, ionizing them, and producing the non-centric CL emission we uncover in our sample (Figure \ref{fig:spatial}; e.g., \citealt{1988MNRAS.235..403T, 2011ApJ...739...69M}). It is also possible that this CL emission is tracing a different species of gas, that is not a CL, from a non-companion galaxy within the MaNGA FoV (at a different redshift than the target galaxy; e.g., J1613 - Figure \ref{fig:spatial}).
\section{Summary and Future Work}
\label{sec:conclusion}
We construct the most extensive sample of MaNGA CL galaxies to date. With our custom pipeline, we measure emission from [NeV]$\lambda 3347$, [NeV]$\lambda 3427$, [FeVII]$\lambda 3586$, [FeVII]$\lambda 3760$, [FeVII]$\lambda 6086$, and/or [FeX]$\lambda$6374 at $\ge$ 5$\sigma$ above the background continuum in 71 galaxies in MaNGA's MPL-11 catalog of 10,010 unique galaxies. 		

Our main findings are:
\begin{enumerate}
	
\item The average size of the CLR for [NeV], [FeVII], and [FeX] is 1.9 kpc, 3.8 kpc, and 2.5 kpc, respectively - beyond the BLR and into the traditional NLR.
	 
\item The fraction of  [NeV], [FeVII], and [FeX]  galaxies with at least one CL-emitting spaxel in their nuclear 2.$^{\prime\prime}$5 region is 98.5\%, 73\%, and 75\%, respectively. Nuclear CL emission is preferentially found in [NeV] galaxies. 


\item We identify two main populations of CL galaxies: 1) galaxies that mostly feature [NeV] emission (33/71 CL galaxies), with relatively high [OIII] and bolometric luminosities (mean [NeV] log([OIII]) luminosity = 41.5 erg s$^{-1}$; mean [NeV] log(L$_{\rm{bol}}$) = 44.5 erg s$^{-1}$), and a high fraction of confirmed AGN (94\%), and 2) galaxies that predominately emit [FeVII] and [FeX] (40/71 CL galaxies), with relatively low [OIII] and bolometric luminosities (mean [FeVII] and [FeX] log([OIII]) luminosities are 40.1 erg s$^{-1}$ and 39.8 erg s$^{-1}$, respectively; mean [FeVII] and [FeX]  log(L$_{\rm{bol}}$) values are 43.7 erg s$^{-1}$ and 43.5 erg s$^{-1}$, respectively), and a low fraction of confirmed AGN (14\% and 25\%, respectively).  

\item 100\% of the [NeV] spaxels in our sample are either BPT AGN or BPT composite, 91\% of the [FeVII] spaxels, and 88.3\% of the [FeX] spaxels. The CLs are strong tracers of BPT AGN and BPT composite sources. 

\item We detect a low number of iron CL galaxies in high E(B - V) value galaxies (E(B - V) $\ge$ 0.045; nine iron CL galaxies) vs. low E(B - V) galaxies (E(B - V) $<$ 0.039; 34 iron CL galaxies). We reason that the destruction of iron CLs by dust grains, which is inversely proportional to AGN bolometric luminosity, may likely be  depleting [FeVII] and [FeX] emission, particularly in the nuclear region where the presence of dust is greater. The [FeVII] and [FeX] galaxies may be tracing lower luminosity AGN, which are possibly too weak to be confirmed by traditional AGN detection techniques.

\item SNR and [OI] shock excitation are viable CL production mechanisms; however, they are not likely primary, as the abundance of SNR and [OI] shocks does not vary significantly across our sample for galaxies with or without: nuclear CL emission, an AGN, or a merging companion. 
\end{enumerate} 
We will explore the CLR kinematics in a future publication to better comprehend the role of outflows on CL production. 
In particular, we will use [OIII] flux maps to evaluate the likelihood that AGN outflows produce CL emission, provided the strong correlation between [OIII] emission and AGN outflows (e.g., \citealt{2017ApJ...835..222S, 2018ApJ...867...66C}). 
Further, we will measure the rotation and cloud velocities of the gas for each CL galaxy (to determine how the bulk motion of gas in the CL galaxies correlates with CL emission), and also analyze the emission line profiles of the CLs to determine if, for example, the CLs feature any blue shifted emission - indicative of outflows. 

Moreover, additional multi-wavelength observations of the CLs would help deduce their nature. X-ray observations from \textit{Chandra}, for example, would allow us to better confirm the population of low luminosity AGN in the CL galaxies. This will help to determine the effectiveness of using CL emission as an unambiguous tracer of AGN in large-scale spectroscopic surveys of galaxies. Finally, our work here is also relevant for motivating near IR measurements of additional CLs that  are observable by the \textit{James Webb Space Telescope}, particularly in cases where optical CLs may be obscured.
\acknowledgments
\newpage
\section*{Acknowledgments}
J.N. and J.M.C. acknowledge support from NSF AST1714503 and NSF AST1847938. 

Funding for the Sloan Digital Sky Survey IV has been provided by the Alfred P. Sloan Foundation, the U.S. Department of Energy Oﬃce of Science, and the Participating Institutions. SDSS-IV acknowledges support and resources from the Center for High-Performance Computing at the University of Utah. The SDSS web site is \href{www.sdss.org}{www.sdss.org}.

SDSS-IV is managed by the Astrophysical Research Consortium for the Participating Institutions of the SDSS Collaboration including the Brazilian Participation Group, the Carnegie Institution for Science,

Carnegie Mellon University, the Chilean Participation Group, the French Participation Group, Harvard Smithsonian Center for Astrophysics, Instituto de Astrof\'{i}sica de Canarias, The Johns Hopkins University, Kavli Institute for the Physics and Mathematics of the Universe (IPMU) / University of Tokyo, the Korean Participation Group, Lawrence Berkeley National Laboratory, Leibniz Institut f\"{u}r Astrophysik Potsdam (AIP), Max-Planck-Institut f\"{u}r Astronomie (MPIA Heidelberg), Max-Planck-Institut f\"{u}r Astrophysik (MPA Garching), Max-Planck-Institut f\"{u}r Extraterrestrische Physik (MPE), National Astronomical Observatories of China, New Mexico State University, New York University, University of Notre Dame, Observat\'{a}rio Nacional / MCTI, The Ohio State University, Pennsylvania State University, Shanghai Astronomical Observatory, United Kingdom Participation Group, Universidad Nacional Aut\'{o}noma de M\'{e}xico, University of Arizona, University of Colorado Boulder, University of Oxford, University of Portsmouth, University of Utah, University of Virginia, University of Washington, University of Wisconsin, Vanderbilt University, and Yale University. 

This publication makes use of data products from the Wide-ﬁeld Infrared Survey Explorer, which is a joint project of the University of California, Los Angeles, and the Jet Propulsion Laboratory/California Institute of Technology, funded by the National Aeronautics and Space Administration.

This research has made use of data supplied by the UK Swift Science Data Centre at the University of Leicester.

This work utilized the Summit supercomputer, which is supported by the National Science Foundation (awards ACI-1532235 and ACI-1532236), the University of Colorado Boulder, and Colorado State University. The Summit supercomputer is a joint effort of the University of Colorado Boulder and Colorado State University.

Software: This work made use of Astropy\footnote{http://www.astropy.org}: a community-developed core Python package and an ecosystem of tools and resources for astronomy \citep{2013A&A...558A..33A, 2018AJ....156..123A, 2022ApJ...935..167A}.



\end{document}